\documentclass[aps,prd,preprintnumbers,showpacs]{revtex4}
\usepackage{graphicx}

\begin{document}

%
%

\eprint{Nisho-1-2024}
\title{Axion Detection with Quantum Hall Effect}
\author{Aiichi Iwazaki}
\affiliation{International Economics and Politics, Nishogakusha University,\\ 
6-16 3-bantyo Chiyoda Tokyo 102-8336, Japan }   
\date{Jan. 14, 2024}
\begin{abstract}
Plateau-plateau transition in integer quantum Hall effect
is a phase transition between metal and insulator. The behavior how the width $\Delta B$ of the transition changes 
with temperature and frequency of radiations imposed has been explored extensively. 
It decreases with the decrease of temperature and frequency, but saturates at critical temperature or frequency.
We have recently discussed\cite{iwa} the effect of axion dark matter on the saturation. 
The axion generates radiations under strong magnetic field in the experiment of quantum Hall effect.
The radiations play a similar role to the one of radiations imposed externally.
In this paper we discuss in detail how the width behaves in temperature and frequency
under the effect of axion dark matter. 
We show that the axion effect can be observable in low temperature roughly below $100$mK.
According to our detailed analysis of the saturation, we find that critical frequency of saturation observed in
previous experiment\cite{doo,doo1} strongly suggests axion mass $m_a=(0.95\sim 0.99)\times 10^{-5}$eV.  
\end{abstract}
\hspace*{0.3cm}

\hspace*{1cm}

\maketitle

\section{introduction}
Finding axion dark matter is one of most significant issues in particle physics. 
It is the important step toward a new physics beyond the standard model of particle physics. It also gives rise to
a solution of dark matter in the Universe.
Axion is the Goldstone boson of Peccei Quinn symmetry\cite{axion1,axion2,axion3}, which naturally solves strong CP problem.
Such an axion is called as QCD axion. The axion mass is severely restricted such as $m_a=10^{-6}\mbox{eV} \sim10^{-3}$ eV
\cite{Wil,Wil1,Wil2}.
In the present paper we only consider QCD axion and 
use physical units, $c=1$, $k_B=1$ and $\hbar=1$. Our result is also applicable to dark photon with frequencies of microwaves
discussed below. 

The QCD axion produces electromagnetic radiations under strong magnetic field $B$. The property is used for the exploration of the axion dark matter
\cite{admx,carrack,haystac,abracadabra,organ,madmax,brass,cast,sumico,iwazaki01}. Some of them 
have been proposed and many of them are undergoing at present\cite{new}.
Obviously, the radiations produced by the axion are also present in the experiments of quantum Hall effect.
It is natural to expect that 
their effects may be observed in some of properties of quantum Hall effect, although they are quite weak.
In particular, as we have shown in previous paper\cite{iwa}, they may appear in much low temperature less than $100$mK.

\vspace{0.1cm}
Quantum Hall effect\cite{von,girvin} is realized in two dimensional electron system under strong magnetic field. 
The system shows various intriguing phenomena such as not only quantization of Hall resistance but also
Josephson-like effect\cite{joseph1,joseph2,joseph3} e.t.c..  Quantum Hall system has been extensively investigated since the discovery, 
but some of phenomena are not still fully understood. One of the phenomena is the saturation\cite{sat1,sat2,wanli,sat3,sat4,sat5,sat6} of the width
$\Delta B$ in plateau-plateau transition of integer quantum Hall effect.
The width defines the range of the magnetic field within which plateau-plateau transition takes place.

\vspace{0.1cm}
Plateau-plateau transition in integer quantum Hall effect is a phase transition between metal and insulator.
It has been extensively explored and the width $\Delta B$ in the transition has been shown to follow
a scaling law such as  $\Delta B \propto T^{\kappa}$ with $\kappa\sim 0.42$\cite{deltaB} as $T\to 0$. 
It has also been shown experimentally that the width saturates in low temperature, 
that is, it never decrease more with the decrease of temperature; $\Delta B=\mbox{const.}$ below a critical temperature.
In general, the saturation is considered to arise owing to finite size effect of two dimensional electrons, 
because the scaling is expected in infinitely large system. 
But recent experiments\cite{sat4,sat5} suggest
that the saturation is caused by not finite size effect, but intrinsic decoherence, although the mechanism of the decoherence is
still unclear. 

Furthermore, similar scaling law holds when we imposed microwaves on Hall bar\cite{engel,balaban,hohls,saeed,doo}.
That is, $\Delta B \propto f^{\kappa}$ with $\kappa=0.4\sim 0.7$ as the frequency $f\to 0$ of the microwaves. 
It has been also observed that the width saturates at a critical frequency $f_s$
when we decrease the frequency of the microwaves just as in the case of the  
temperature.
( In actual experiments we use AC voltage for the measurement of the width. )

\vspace{0.1cm}
In our previous paper\cite{iwa}, we have shown a possibility that the saturation arises owing to the effect of the axion dark matter.
Especially, electromagnetic radiations ( actually, microwaves ) generated by the axion cause the decoherence
of electrons in low temperature. Using the analysis, we have proposed a way of axion detection using microwaves imposed on Hall bar.
In this paper, we discuss in detail how critical temperature and frequency of the microwaves at which the saturation arises,
depends on the physical parameters like size of Hall bar, temperature and axion mass. 
According to the analysis, 
we propose more detailed way of
the axion detection in the present paper than previous one. In particular, we 
present two conditions which critical frequency $f_s$ must satisfy in order to give the axion mass, i.e. $f_s=m_a/2\pi$. 
One is that $f_s$ does not depend on temperature and the other one is that 
$f_s$ does not depend on the size of Hall bar.  
Here we would like to point out that 
the previous experiment\cite{doo,doo1} using microwaves strongly suggests axion mass $m_a\simeq 0.95\times 10^{-5}$eV,
because it seems that the critical frequency observed in the experiment satisfies two conditions presented in this paper.

\vspace{0.1cm}
Here we naively explain how the axion dark matter affects on plateau-plateau transition.
We consider infinitely large two dimensional electrons at zero temperature under strong magnetic field. Owing to disorder potential,
almost all electrons in each Landau levels occupy localized states except for electrons occupying extended state with energy $E_c$
located in the center of each Landau level. As magnetic field $B$ decreases, Fermi energy $E_f$ increases. As long as the Fermi energy is less than
$E_c$, Hall conductivity stays in a plateau.
When Fermi energy passes the energy $E_c$ of extended states, Hall conductivity goes to next plateau so that 
the plateau-plateau transition looks like a step function. It is very sharp, i.e. $\Delta B=0$. On the other hand, the axion dark matter 
generates radiations under the strong magnetic field so that the radiations are absorbed by electrons.
Even if Fermi energy $E_f$ is below the energy $E_c$, some of localized electrons transit to the states with energies
larger than $E_c$ by absorbing the radiations. These electrons loose their energies by emitting phonons and
may occupy the extended states with the energy $E_c$. Thus, there is non zero probability of
the extended state occupied even if $E_f<E_c$. Therefore, the plateau-plateau transition becomes smooth function
of magnetic field, i.e. $\Delta B\neq 0$. The smooth transition $\Delta B\neq 0$ is not finite size effect.
This is a naive explanation how the axion dark matter causes visible effect on the plateau-plateau transition.

We should mention that although thermal effect at $T\neq 0$ contributes to the transition, the axion effect dominates over the thermal effect
as long as temperature $T$ is very low. Indeed, we will show later that the axion effect is dominant
approximately for $T<100$mK.



\vspace{0.1cm}
In this paper
we also propose a way of confirmation that the axion dark matter really causes the saturation in temperature or frequency of
microwave imposed. Using parallel conducting slabs which sandwiches Hall bar, we shield radiations by the axion
so that we expect the absence of the saturation of the width $\Delta B$. 
 ( Sometimes in literatures, the derivative $d\rho_{xy}/dB$ at the center in the plateau transition instead of $\Delta B$ 
 is used to see how it behaves with temperature or frequency of radiation.  
The saturation of $\Delta B$ corresponds to
the saturation of $d\rho_{xy}/dB$. The decrease of $\Delta B$ corresponds to the increase of $d\rho_{xy}/dB$. )

\vspace{0.1cm}
In the next section (\ref{2}), we briefly explain energy scale in integer quantum Hall effect and our notation used in the paper. We proceed 
to explain localization of electrons in quantum Hall effect in section (\ref{3}).
Most of electrons occupy localized states but a small fraction of electrons occupy extended states,
which may carry electric current. 
The localization leads to plateau of Hall conductivity or resistivity in quantum Hall effect.
In this section, using energy distribution of electrons at zero temperature, we discuss how Hall conductivity
behaves and forms plateau according to the variation of magnetic field $B$. 
We define width $\Delta B$ in 
plateau-plateau transition used in the paper.
The dependence of the width on temperature or axion mass is discussed in later sections. 
In the section (\ref{4}), we briefly explain the axion dark matter. We consider QCD axion as dark matter candidate.
We show how the axion 
generates electromagnetic radiations with energy $m_a$ under external magnetic field, which is used in experiment of quantum Hall effect. 
In the next section (\ref{5}),
we discuss in detail how the width $\Delta B$ depends on temperature and axion mass. 
We find that the saturation of $\Delta B$ in temperature
only arises in the presence of the axion dark matter. 
Without the axion effect, the saturation does not appear even in the system with finite size.
In the next section (\ref{6}), we discuss the effect of external microwaves on the plateau-plateau transition.  
We show how the width $\Delta B$ depends on frequency $f$, temperature and axion mass.
Especially, we show the presence of critical frequency $f_s$ below which the width $\Delta B$ does not depend on the frequency $f$ ( $<f_s$ ).
We present two conditions which the saturation frequency $f_s=m_a/2\pi$ must satisfy to give the axion mass. 
In the section ( \ref{7} ), we estimate energy power generated by axion dark matter in two dimensional electrons.
We compare it with thermal noise and find that the axion effect can be observable in low temperature less than $100$mK, at least
for the surface area of two dimensional electrons being large such as $10^{-3}\rm cm^{2}$ or larger. 
The axion effect becomes larger as the size of Hall bar becomes larger, because the energy of the axion received by electrons is
bigger as the number of electrons becomes larger.
In the final section ( \ref{8} ), we propose a way how we confirm the presence of the axion effect
in quantum Hall effect.  The point is that we shield the radiations generated by the axion dark matter by
using conducting slabs put around Hall bar.

\section{energy scale in integer quantum hall effect}
\label{2}
We briefly explain energy scales relevant to quantum Hall effect\cite{girvin}.
The quantum Hall effect is realized in two dimensional electrons of semiconductors under magnetic field $B$. 
The states of the two dimensional free electrons are specified by integer $n \geq 0$, so called Landau levels.
There are a number of degenerate states in each Landau level
with the degeneracy $eB/2\pi$ ( i.e. number density of degenerate states in a Landau level ). 
The typical scale of the magnetic field is of the order of $10$T.
Each electrons oscillate with cyclotron frequency $\omega_c=eB/m^{\ast}$ where mass $m^{\ast}$ denotes effective one of electron
in semiconductors. Generally $m^{\ast}$ is much smaller than real mass $m_e\simeq 0.51$MeV of electron, e.g.
$m^{\ast}=0.067m_e$ in GaAs. Then, cyclotron frequency ( energy ) $\omega_c$ is of the order of $\sim 10^{-2}(B/10T)$eV.
Furthermore, their energies are specified with integer $n \ge 0$ such that 
$E_n=\omega_c( n+1/2)$. The wave functions are extended with typical length scale,
so called magnetic length $l_B=\sqrt{1/eB}$. 
$l_B\simeq 8.2\times 10^{-7}\rm cm \sqrt{(10T/B)}$.
It is cyclotron radius of electron under the magnetic field $B$.

\vspace{0.1cm}
Electron possesses spin components with up and down so that
each Landau level is split to two states with energies $E_{n\pm}=\omega_c(n+1/2)\pm g\mu_B B$ owing to Zeeman effect.
Here we note that $g\simeq 0.44$ and Bohr magneton $\mu_B=e/2m_e$.
Zeeman energy is of the order of $10^{-3}(B/10T)$eV.  
It is smaller than the cyclotron energy $\omega_c$.

We mainly consider axion mass $m_a$ in a region $10^{-5}\mbox{eV}\ge m_a \ge 10^{-6}\mbox{eV}$ so that
the mass is smaller than Zeeman energy and cyclotron energy $\omega_c$.
The mass is of the order of or less than the width ( extension ) $\Delta E$ in Fig.\ref{a} of 
the energy distribution of electrons in a Landau level, as we will
explain in next section. Therefore, because the energies of radiations produced by the axion are almost identical to $m_a$,
the effect of the radiations can be observable because electrons may absorb the radiations within a Landau level.
 
We use an index so called filling factor $\nu\equiv \rho_e/(eB/2\pi)=2\pi\rho_e/eB$ to specify which Landau levels are occupied;
$\rho_e$ denotes number density of electrons
( typically, $\rho_e\sim 10^{11}/\rm cm^2$ ). For instance,
it implies Landau level with energy $E_{0-}$ is fully occupied but Landau level with $E_{0+}$ is
partially occupied when $2 > \nu >1$.

\vspace{0.1cm}

It is remarkable feature\cite{von,aokiando,halperin} of quantum Hall effect that Hall resistance ( conductance ) is quantized such that 
$\rho_{xy}=(2\pi/e^2)\times 1/n$ ( $\sigma_{xy}=n\times e^2/2\pi$ ) with positive integer $n$ specifying Landau level. It 
is constant within a range $n+1>\nu>n$ when we vary the magnetic field $B$; $\nu=2\pi\rho_e/eB$.
In the range, Landau levels up to $n$ are fully occupied, while
the level with $n+1$ is partially occupied. That is, we see plateaus in the diagram of $\rho_{xy}$ ( $\sigma_{xy}$ ) in $B$. 
The plateaus arise owing to the localization of two dimensional electrons discussed soon below. 
The localization arises owing to disorder potential for electrons. That is, almost of electrons are trapped in the potential.

\section{localization of two dimensional electrons}
\label{3}
\begin{figure}[htp]
\centering
\includegraphics[width=0.6\hsize]{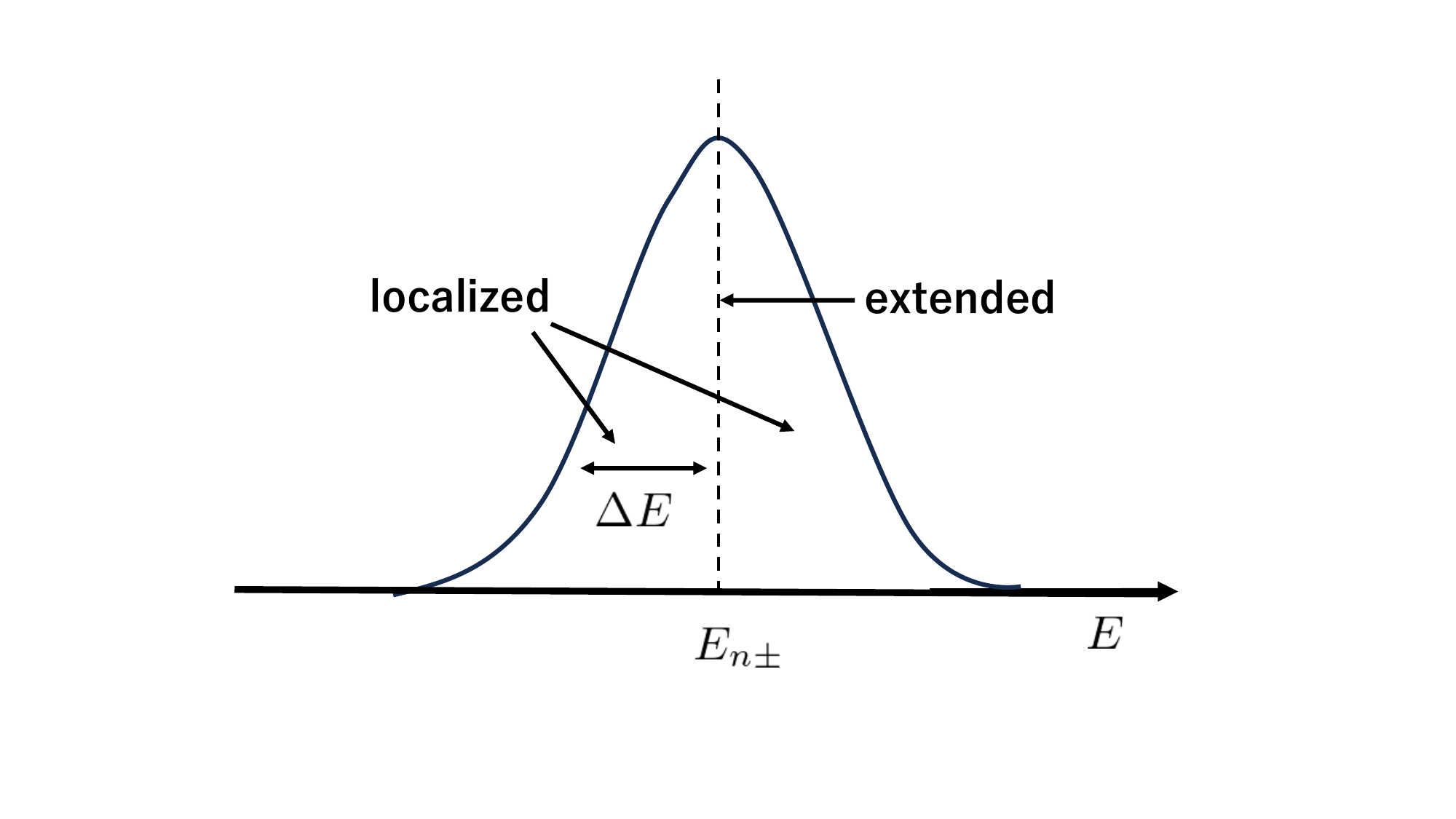}
\caption{Density of state $\rho(E)$. Dashed line denotes $\rho(E)\propto \delta(E-E_{n\pm})$ without the effect of random potential $V$.}
\label{a}
\end{figure}

We explain localization of two dimensional electrons under strong magnetic field $B$.
In the case of free electrons we have the density of state $\rho(E)\propto \sum_{n=0,1,,,} \delta (E-E_{n\pm})$.
But,
there are impurities, defects e.t.c. in actual materials. They lift up the degeneracy in Landau level. 
Electrons are severely affected by a disorder potential $V$.
Most of electrons are localized and they
cannot carry electric currents. But, a small fraction of them are not localized so that they can carry electric current.
It means that Hall resistance only receives the effect of non-localized electrons. Localized electrons do not contribute
to Hall resistance.
The essence in integer quantum Hall effect is 
the presence of non-localized ( extended )\cite{aokiando,ono} states of electrons under 
strong magnetic field $B$.
Although the disorder potential $V$ localizes almost of all electrons, there exist a non-localized state with energy $E=E_{n\pm}$.
According to numerical simulations we find that 
in the presence of a potential $V$, 
the density of states $\rho(E)$ has finite width around the energy $E_{n\pm}$
shown schematically in Fig.\ref{a}. In the figure we show localized states and extended states.
Extended localized state is located at $E=E_{n\pm}$.

The density of state $\rho(E)$ has finite width $\Delta E$ owing to the potential $V$.
It is generally supposed that the width is less than the cyclotron frequency $\omega_c=eB/m^{\ast}$ in strong magnetic field $B$.
That is, the potential energy is much smaller than the cyclotron frequency:
$\Delta E \ll \omega_c$ ( $\omega_c\sim 10^{-2}$eV with $B\sim 10$T ). 
For instance, $\rho(E)\propto \sqrt{1-((E-E_{n\pm})/\Delta E)^2}$ with $|E-E_{n\pm}| \ge \Delta E$\,\cite{andouemura}. 
Furthermore, it is supposed that the potential has
the same amount of attractive and repulsive components. So, the form of $\rho(E)$ is symmetric around the center
$E_{n\pm}$ as shown in Fig.\ref{a}.

\vspace{0.1cm}
We may define coherent length as the size of localized state. 
It depends on the energy of the state. Then,
the references\cite{aoki1,aoki2} have shown in the system with infinitely large size that
a coherent length $\xi(E)$ diverges such as $\xi(E)\to |E-E_{n\pm}|^{-\nu}$
as $E\to E_{n\pm}$ with $\nu\sim 2.4$. 
It implies that there are extended states with their energy $E_{n\pm}$. 

\vspace{0.1cm}
When Fermi energy $E_f$ is less than $E_{n+}$ but larger than $E_{n-}$, the energies of electrons occupying
extended states are equal to $E_{n-}$ or $E_{m\pm}$ with $n>m$. 
For instance, for $E_{1+}>E_f>E_{1-}$,
Hall resistance is given such that
$\rho_{xy}=(2\pi/e^2)\times 1/3$.
As long as the Fermi energy $E_f$ is in the range $E_{1+}>E_f>E_{1-}$, the Hall resistance does not vary
with magnetic field $B$. A plateau is formed.
On the plateau, we have vanishing longitudinal resistance $\rho_{xx}=0$.
Only electrons in extended states
with energies such as $E_{1-}$ and $E_{0\pm}$ carry electric currents. 
When the electron occupies the extended states with the energy $E_f=E_{1+}$, it carry electric current. Then, Hall resistance
suddenly down to $ \rho_{xy}=(2\pi/e^2)\times 1/4$, or Hall conductivity rises up to $\sigma_{xy}=4\times e^2/2\pi$
from $\sigma_{xy}=3\times e^2/2\pi$. The transition is sharp like the step function.
Hereafter, we mainly state Hall conductivity because it is easy to see the effect on the conductivity of 
electrons carrying electric currents. Generally, electric conductivity is proportional to electron density
carrying electric current.

It should be noticed that even if a single electron occupies the extended states with the energy $E_{1+}$, 
the Hall conductance jumps to the next plateau, for instance, $\sigma_{xy}=4\times e^2/2\pi$ from $\sigma_{xy}=3\times e^2/2\pi$.
This is a striking feature of quantum Hall effect. The feature is understood in topological argument\cite{topology1,topology2}.
Hereafter, we examine in detail the case of $E_{1-}< E_f <E_{1+}$ for concreteness.

\vspace{0.1cm}
The above argument only holds in Hall bar with the infinite large size. 
Extended states only have the energies $E_{n\pm}$. In actual Hall bar with finite size,
localized states are present whose sizes are larger than 
the size of Hall bar. Because of the divergence of coherent length $\xi(E)$
as $E$ approaches $E_{n\pm}$, we understand the presence of such localized states with their sizes larger than Hall bar.
Electrons in such states have energies $E$ in the range $E_{n\pm}-\delta \le E \le E_{n\pm}+\delta$
where $\delta=\delta(L_h)$ depends on the size $L_h$ of Hall bar such as $\delta (L_h\to \infty)=0$.
In general $\delta$ is smaller than the width $\Delta E$ in $\rho(E)$. 
We may call such states as extended states because electrons in the states can carry electric current.

\begin{figure}[htp]
\centering
\includegraphics[width=0.8\hsize]{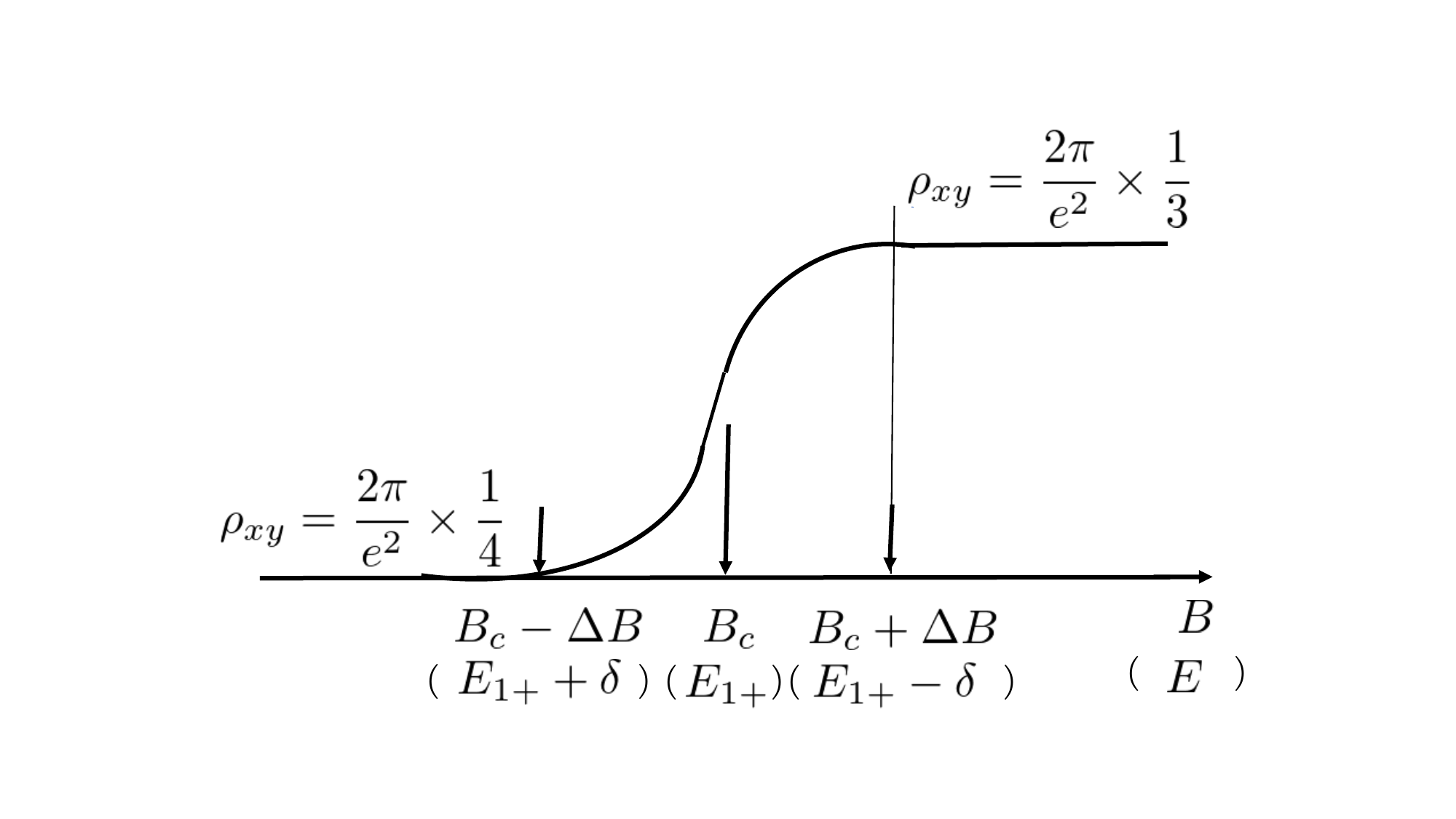}
\caption{Plateau-plateau transition. We define $\Delta B$ in the figure. }
\label{aa}
\end{figure}

\vspace{0.1cm}
In Hall bar with finite size, plateau-plateau transition takes place smoothly. When Fermi energy $E_f$ increases, but stay less than $E_f=E_{1+}-\delta$,
Hall conductivity stays in a plateau. However, when it reaches at the energy $E_f=E_{1+}-\delta$,
the transition begins and the conductivity increases smoothly as $E_f$ increases. $B$ takes the value of $B_c+\Delta B$ at $E_f=E_{1+}-\delta$.
When $E_f=E_{1+}$, the magnetic field is 
given by $B_c=2\pi\rho/4e$.
Eventually when $E_f$ reaches at $E_{1+}+\delta$,
the conductivity stops to increase and stay at next plateau. $B$ takes the value of $B_c-\Delta B$ at $E_f=E_{1+}+\delta$.
Thus, we have $\Delta B\neq 0$.
The plateau-plateau transition mentioned here is the one in Hall bar with finite size at 
zero temperature $T=0$. See Fig.\ref{aa}.
In the figure, we define the width $\Delta B$ between two plateaus.
The Hall resistance begins to decrease at $E_f=E_{1+}-\delta$ and the decrease stops at $E_f=E_{1+}+\delta $.
In our discussion we use the width $\Delta B$ as specified in the figure.

\section{axion dark matter}
\label{4}
The axion is a Nambu Goldstone boson associated with Peccei Quinn symmetry
introduced  for solving strong CP problem. 
The symmetry is imposed in a model beyond the standard model, but it is spontaneously broken. As a result, the axion
appears.
The axion as the boson $a(t,\vec{x})$\cite{axion1,axion2,axion3} acquire its mass $m_a$ due to instanton effects in QCD. Because the axion 
couples with electromagnetic fields with coupling $g_{a\gamma\gamma}$, Maxwell equations are modified\cite{iwazaki01} in the following,

\begin{eqnarray}
\label{modified}
\vec{\partial}\cdot\vec{E}+g_{a\gamma\gamma}\vec{\partial}\cdot(a(t,\vec{x})\vec{B})&=0&, \quad 
\vec{\partial}\times \Big(\vec{B}-g_{a\gamma\gamma}a(t,\vec{x})\vec{E}\Big)-
\partial_t\Big(\vec{E}+g_{a\gamma\gamma}a(t,\vec{x})\vec{B}\Big)=0,  \nonumber  \\
\vec{\partial}\cdot\vec{B}&=0&, \quad \vec{\partial}\times \vec{E}+\partial_t \vec{B}=0.
\end{eqnarray}
with electric $\vec{E}$ and magnetic $\vec{B}$ fields,
where we take the axion field $a(t,\vec{x})$ as the one representing the axion dark matter.

From the equations, we obtain the electric field $\vec{E}_a$
generated by the axion $a(t,\vec{x})$  
under static external magnetic field $\vec{B}$. Namely, when magnetic field $\vec{B}$
is present, the axion dark matter generates oscillating electric field $\vec{E}_a$.
Because the parameter $g_{a\gamma\gamma}a(t,\vec{x})$ is extremely small as shown soon below, we obtain

\begin{equation}
\label{E}
\vec{E}_a(t,\vec{x})=-g_{a\gamma\gamma}a(t,\vec{x})\vec{B}.
\end{equation}
with $g_{a\gamma\gamma}=g_{\gamma}\alpha/f_a\pi$,
where $\alpha\simeq 1/137$ denotes fine structure constant and $f_a$ is axion decay constant
satisfying the relation $m_af_a\simeq 6\times 10^{-6}\rm eV\times 10^{12}$GeV in the QCD axion.
The parameter $g_{\gamma}$ depends on the axion model, i.e. 
$g_{\gamma}\simeq 0.37$ for DFSZ model\cite{dfsz,dfsz1} and $g_{\gamma}\simeq -0.96$ for KSVZ model\cite{ksvz,ksvz1}.
The mass of the QCD axion is severely restricted such as $m_a=10^{-6}\mbox{eV} \sim10^{-3}$ eV,
\cite{Wil,Wil1,Wil2}. In the present paper we mainly consider the mass $10^{-5}\mbox{eV}\ge m_a\ge 10^{-6}\mbox{eV}$.
We should mention that the parameter $g_{a\gamma\gamma}$ is automatically determined in QCD axion
when we take the value of the axion mass, i.e. $f_a\simeq (6\times 10^{-6}\mbox{eV}/m_a)\times 10^{12}$GeV.

\vspace{0.1cm}
The amplitude of the axion dark matter $a(t,\vec{x})\simeq a_0\cos(m_at)$ is extremely small. 
( The momentum of the axion dark matter is of the order of $10^{-3}m_a$ 
so that we may neglect the momentum. )
Supposing that the dark matter in the Universe is composed of the axion, we find that the local energy density $\rho_d$ of the dark matter
is given as
$\rho_d=m_a^2\overline{a(t,\vec{x})^2}=m_a^2a_0^2/2\sim 0.3\rm GeV/cm^3$; $\overline{Q}$ denotes
time average of the quantity $Q$. Then we find that $g_{a\gamma\gamma}a(t,\vec{x})\sim 10^{-21}$.
Although the electric field $\vec{E}_a=-g_{a\gamma\gamma}a(t,\vec{x})\vec{B}$
\cite{sikivie, iwazaki01} is extremely small, 
it is inevitably produced 
in the experiment of quantum Hall effect because of the presence of magnetic field $B \sim 10$T.
This oscillating electric field $\vec{E}_a$ generates electromagnetic radiations\cite{iwazaki01} from conductors.
It makes electrons in metals oscillate so that the oscillating electrons emit electromagnetic radiations.
Indeed, Hall bar is surrounded by metals composing mixing chamber for cooling the Hall bar, superconducting magnet e.t.c.
The frequency $f$ ( wave length ) of the radiations is given by the axion mass, $f=m_a/2\pi$ ( $m_a^{-1}$ ).
Such radiations are absorbed by electrons in Hall bar and they affect on the width $\Delta B$ in plateau-plateau transition,
which is our main concerns in the paper.

We mainly focus on the mass $m_a$ such as  $10^{-5}\mbox{eV}\ge m_a > 4\times10^{-6} $eV 
( the corresponding frequency $m_a/2\pi$ of radiation is about $1\mbox{GHz} \sim 2.4\mbox{GHz}$. 
The wave length $12\mbox{cm}\sim 30$cm is much larger than typical size of Hall bar. )
We will find later that axion mass $m_a=(0.95\sim 0.99)\times 10^{-5}$eV, by analyzing a previous experiment\cite{doo,doo1} 
using microwaves imposed in quantum Hall effect.

\section{axion and plateau-plateau transition}
\label{5}

We explain in detail how the width $\Delta B$ depends on axion mass and temperature.
First we suppose the system at zero temperature. Then, the energy distribution of electrons has sharp boundary at Fermi energy $E_f$
when axion effect is neglected.
That is, the states with energies less than $E_f$ are fully occupied and the states with energies larger than $E_f$ are empity.
Thus, when Fermi energy is less than $E_{1+}-\delta$, electric current does not flow and the conductivity $\sigma_{xy}$ stays in 
the plateau, $\sigma_{xy}=3e^2/2\pi$. Hall resistivity takes the value $\rho_{xy}=1/\sigma_{xy}=2\pi/3e^2$. 
When external magnetic field $B$ decreases, Fermi energy increases and reaches at the value $E_{1+}-\delta$.
Then, electric current begins to flow because extended states with energies larger than $E_{1+}-\delta$ begin to be occupied.
The conductivity $\sigma_{xy}$ begins to increase or Hall resistivity decreases.
As $B$ decreases more, Fermi energy increases. When the Fermi energy $E_f$ goes beyond $E_{1+}+\delta$, the conductivity takes the value
$4e^2/2\pi$ and stay in the next plateau, or Hall resistivity takes $2\pi/4e^2$.
Thus, the width $\Delta B$ is determined by the magnetic field $B_c\pm \Delta B$
at which $E_f=E_{1+}\mp \delta$. The width $\Delta E_f$ between $E_f=E_{1+}-\delta $ and $E_f=E_{1+}+\delta $,
leading to $\Delta B$ is

\begin{equation}
\Delta E_f=E_f+\delta - (E_f-\delta )=2\delta \quad \mbox{at temperature}\,\,T=0 \,\,\, \mbox{with no axion effect}
\end{equation}
We see the plateau-plateau transition of Hall resistance
schematically shown in Fig.\ref{aa}.

\begin{figure}
\begin{tabular}{cc}
\begin{minipage}{0.5\textwidth}
\centering
\includegraphics[width=1.0\hsize]{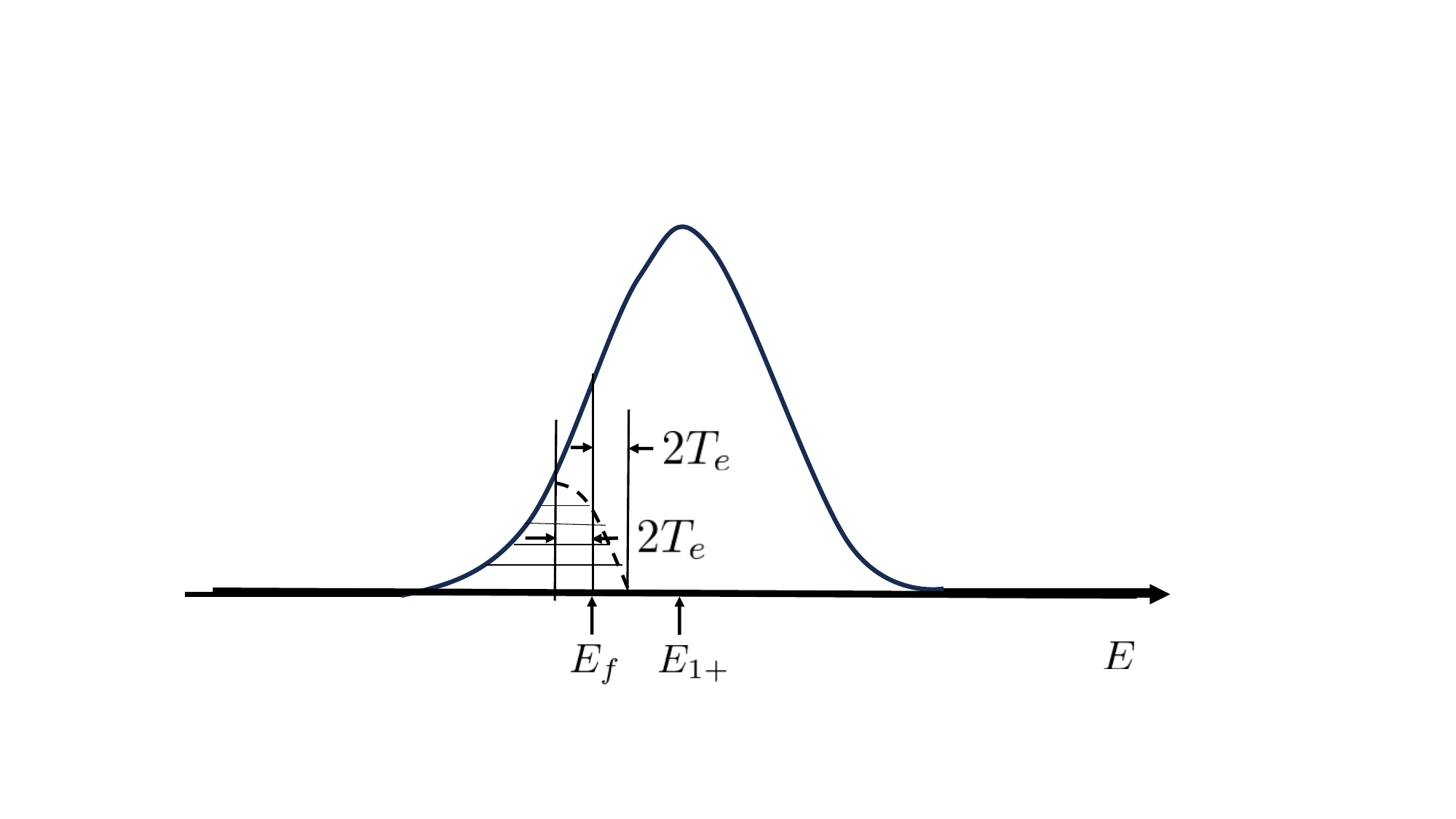}
\caption{Energy distribution of electrons, $E_f+2T_e <E_{1+}$}
\label{bb}
\end{minipage}
\begin{minipage}{0.5\textwidth}
\centering
\includegraphics[width=1.0\hsize]{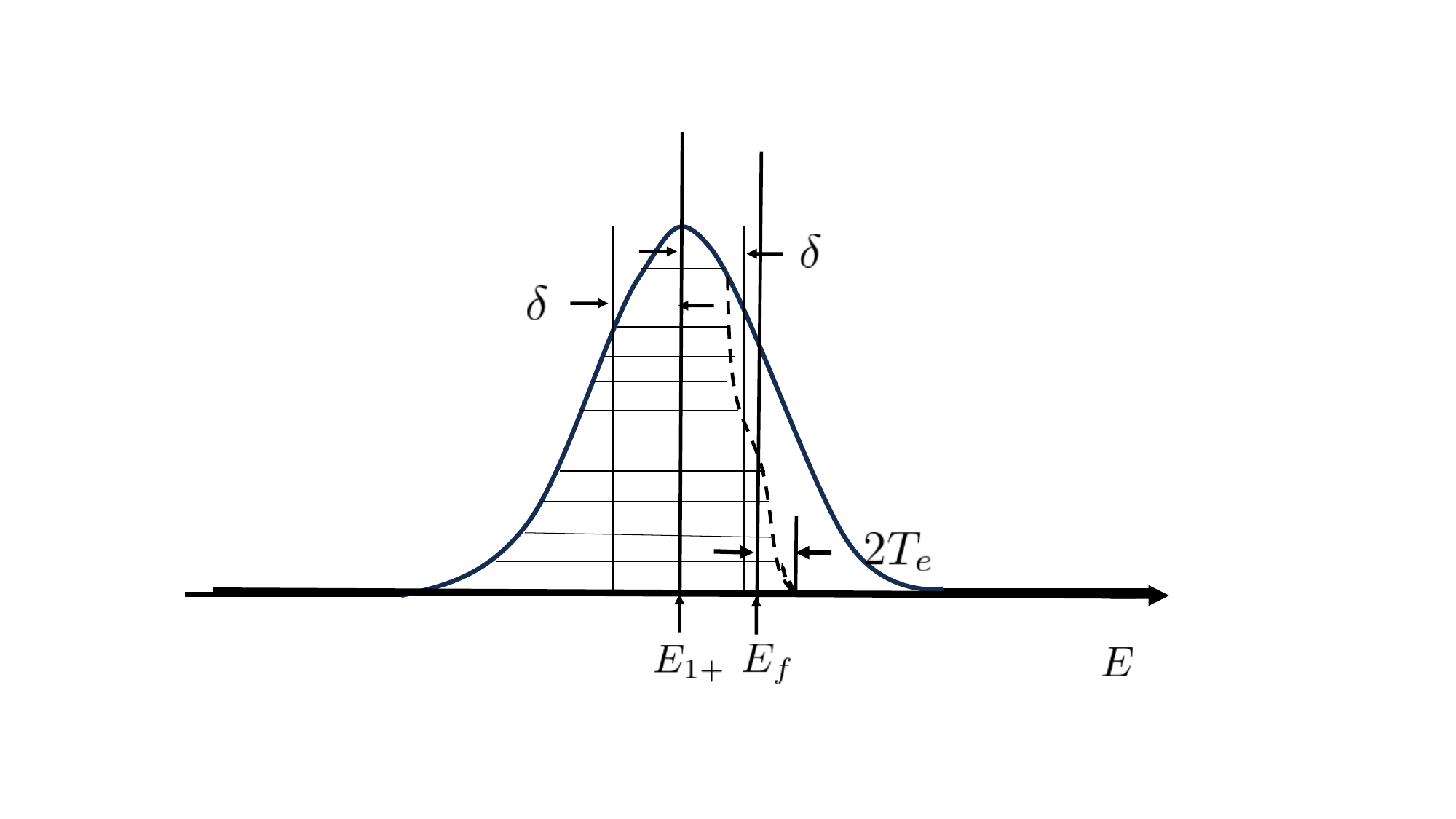}
\caption{Energy distribution of electrons, $E_f>E_{1+}+\delta$ }
\label{cc}
\end{minipage}
\end{tabular}
\end{figure}

\vspace{0.1cm}
When the temperature $T\neq 0$, the energy distribution of electrons has no sharp boundary at chemical potential $\mu$. 
The boundary is smeared out by thermal effect around chemical potential $\mu$. Because we only consider low temperature $ <1$K,
we approximately set $\mu=E_f$.  Although the distribution has no sharp boundary, we may approximately define 
the effective temperature $T_e$ such that the distribution has sharp boundary at $E_f+2T_e$.
Electrons occupy the states below the energy  $E_f+2T_e$ and the states with energies larger than $E_f+2T_e$ are empty.
Especially, the distribution is smeared out over the width $2T_e$ schematically shown in Fig.\ref{bb}.
Such a simplification arises from the fact that the real energy distribution of electrons decreases exponentially $\propto \exp(-(E-E_f)/T)$
for $E>E_f$. Owing to 
the simplification, we can easily understand how the width $\Delta B$ depends on temperature. 
We should note that the temperature $T_e$ is not real temperature $T$ and it depends on the density of state $\rho(E)$.
Thus, when we set $T_e=gT$, the constant $g$ depends on real density of state $\rho(E)$. We expect that $g$ is of the order of $1$.

\vspace{0.1cm}
Then, the width $\Delta B$ is determined in the following. That is, as $B$ decreases, $E_f$ increases. When $E_f+2T_e$ is equal to $E_{1+}-\delta $,
electric currents begin to flow. 
That is the point at which the conductivity begins to increase, i.e.
$B=B_c+\Delta B$.
As $B$ decreases more, Fermi energy $E_f$ passes the energy $E_{1+}$ and eventually, $E_f$ reaches at $E_{1+}+\delta+2T_e $,
see Fig.\ref{cc}.
At the point, the conductivity
stops to increase and stay in next plateau. It is the point $B=B_c-\Delta B$. In other words, Fermi energy moves from $E_f=E_{1+}-\delta-2T_e$
to $E_f=E_{1+}+\delta+2T_e$ in the plateau-plateau transition. The width is given by 

\begin{equation}
\Delta E_f=2\delta +4T_e \quad \mbox{at temperature}\,\,T_e\neq 0 \,\,\,\mbox{with no axion effect} 
\end{equation}

Obviously, the width $\Delta B$ corresponding to $\Delta E_f$ decreases as temperature $T_e$ ( or $T$ ) decreases. 
It takes the value $2\delta $ when $T_e=0$. It never saturate at non zero temperature.
We should make a comment that the dependence of $\Delta B$ on $\Delta E_f$ or temperature $T_s$ is expected 
such that $\Delta B \propto |T_s|^{\kappa}$ with $\kappa \sim 0.43$ \cite{deltaB} as $T_s\to 0$
according to the scaling analysis,
at least in the case of infinitely large Hall bar, i.e. $\delta=0$. Thus, $\Delta B$ is such as $\Delta B\propto (\Delta E_f)^{\kappa}$
as $T_e\to 0$ in the infinitely large Hall bar.

\vspace{0.1cm}

In addition to the decrease of the width $\Delta B$ with temperature, the width $\Delta B$ decreases as size of Hall bar increases.
That is,   $\Delta B$ decreases with $\delta$, which decreases as the size of Hall bar increases. The phenomena have been well known.

\vspace{0.1cm}

Here we make a comment of the effect of our simplification of sharp boundary 
at $E_f+2T_e$ in energy distribution of electrons. Even if we do not take such a simplification, 
there is no critical temperature $T_s$ of saturation. That is, the width $\Delta E_f$ defining the difference between
a starting point and its ending point of the increase of $\sigma_{xy}$, smoothly decreases as real temperature decreases. 
This is because the effect of the temperature $T$ on the energy distribution smoothly decreases as $T$ decreases. 
It never arise that the width does not decrease below a non zero temperature.
Therefore, the presence of the saturation cannot be explained simply by finite size effect.

\vspace{0.1cm}
Here we explain thermal effect on $\delta$. 
Although $\delta$ depends on length scale $L_{h}$ of Hall bar, 
effectively extended states in which electrons carry electric current are not restricted to the energy region such as $E_f+\delta >E>E_f-\delta$ at non zero
temperature.
For sufficiently low temperature, phase coherent length is much larger than the scale 
$L_{h}$ of Hall bar. The phase coherence length\cite{doo1} is the length scale within which quantum states hold keeping
quantum coherence. As temperature increases, the phase coherent length decreases and eventually reaches at the physical size $L_h$ of Hall bar.
There is a critical temperature $T_c$ such that
at the temperature $T_c$, the phase coherent length is equal to $L_{h}$. Beyond the temperature, the coherent length becomes smaller than 
the physical size of Hall bar. It implies that actual extension of localized states diminishes up to the phase coherent length.
It apparently seems that electric current does not flow
because all of localized states have extension less than the size of Hall bar.
But, electrons in the localized states with spatial extension less than $L_h$ may make hopping to  
to nearby localized states and carry electric current. The hopping arises owing to scattering with phonon or impurities 
or tunneling under external electric field. It does not arise between localized states with large energy difference among them.
In this way, electrons in the localized states even with their energies less than $E_{1+}-\delta$
may carry electric currents at temperature $T_e>T_c$. 
As temperature increases more, the phase coherent length decreases more 
and the hopping of electrons gets more actively. Thus, the energies of effectively extended states becomes smaller than $E_{1+}-\delta$.
It implies that we have effective width $\delta_e(T_e)$ instead of real $\delta $ such that $\delta_e(T_e)\ge \delta$ for $T_e \ge T_c$
and $\delta_e(T_e)= \delta$ for $T_e \le T_c$. $\delta_e(T_e)$ increases as $T_e$ increases.
Thus, the effective extended states at $T_e>T_c$  are defined as those with energies $E$

\begin{equation}
E_{1+}+\delta_e(T_e) \ge E \ge E_{1+}-\delta_e(T_e) \quad \mbox{for} \quad T_e\ge T_c
\end{equation}
with $T_c$ given such as $\delta_e(T_c)=\delta$.

\vspace{0.1cm}
In the above argument, we need to replace $\delta$ by the effective width $\delta_e(T)$ or $\delta_e(T_e)$,

\begin{equation}
\Delta E_f=2\delta_e(T_e)+4T_e \quad \mbox{at temperature}\,\,T_e\neq 0 \,\,\,\mbox{with no axion effect}. 
\end{equation}
It decreases smoothly as $T_e$ decreases. It takes the value $\Delta E_f=2\delta+4T_e$ below $T_e\le T_c$
and reaches $2\delta$ as $T_e\to0$.
Therefore,
we find that $\Delta E_f$ never saturate for any temperature $T_e$ when axion effect is absent.
The width $\Delta B$ corresponding to $\Delta E_f$  also does not saturate. It behaves such as $\Delta B=\mbox{const.}+c\,T_e$
as $T_e \to 0$ with a numerical constant $c$.

\begin{figure}[htp]
\centering
\includegraphics[width=0.8\hsize]{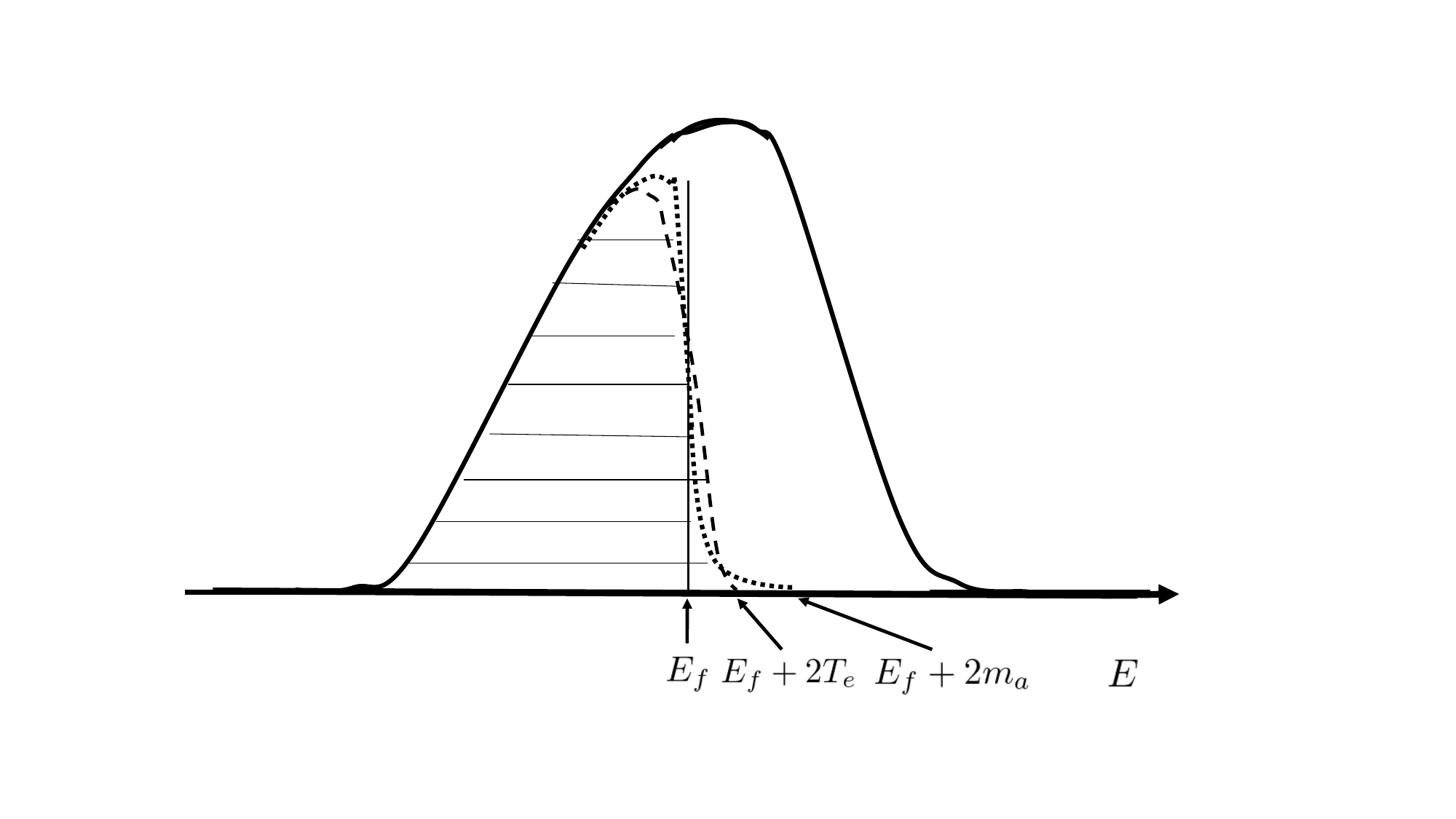}
\caption{Energy distribution smeared by only temperature (dashed) and axion effect added (dotted) 
when $m_a>T_e$}
\label{dd}
\end{figure}

\vspace{0.1cm}
Now we take into account the axion effect.
The dark matter axions produces radiations with energy $m_a$ under strong magnetic field.
Such radiations with their energy $m_a$ are absorbed by electrons, 
which make electrons transit from the state with energy $E$ to the state with energy $E+m_a$.
The electrons in the state with energy $E+m_a$ may loose their energies by emitting phonon.
This is similar process to the one that electron with energy $E$ transits to the state with energy $E+T$ by
absorbing thermal energy $T$ and it loose its energy by emitting phonon. 
( We note that black body radiation with energy $T$ is absorbed by electron just as radiation generated by axion. )
Consequently,  
the energy distribution of electrons is smeared out. But, the effect only slightly modifies the energy distribution because the axion effect is
quite small. It is not important for later discussion to specify precisely how large the axion effect smears the energy distribution 
when temperature is low sufficiently for axion effect to be dominant over thermal effect.
 We only need to know 
the energy at which the energy distribution has sharp boundary. 
Thus, we simplify the effect of the axion such that the energy distribution smeared by axion also has a sharp boundary in the following.
When $T_e>m_a$, the sharp boundary caused by thermal effect
does not change even if the axion effect is taken into account. 
There are no occupied states with energies larger than $E_f+2T_e$ ( $>E_f+2m_a$ ).
On the other hand, when $T_e<m_a$, sharp boundary moves upward beyond $E_f+2T_e$. That is, 
there are occupied states with energies larger than $ E_f+2T_e$, while states with energies larger than $E_f+2m_a$ are empty.
Consequently, the sharp boundary is located at $E_f+2m_a$ ( $>E_f+2T_e$ ) when $m_a>T_e$, while 
at $E_f+2T_e$ when  $m_a<T_e$. See Fig.\ref{dd}.
Similarly to the case of temperature $T_e$, the axion mass $m_a$ in the energy distribution of electrons is not necessarily equal to real axion mass. 
But, when we discuss the effect of microwave imposed externally on Hall bar in later section, the microwave 
plays the similar role to the radiation by the axion. Energy distribution of electrons modified by the radiation 
is supposed to have sharp boundary associated with the frequency $f$ of the radiation. Both radiations by axion and external setup
are supposed to cause sharp boundary on the energy distribution such as the boundary located
at $E_f+2m_a$ or $E_f+4\pi f$ when $m_a>T_e$ or $2\pi f>T_e$. 
In this case we may directly compare the frequency $f$ with axion mass as real ones.

\vspace{0.1cm}
Later we estimate how the axion effect is large compared with thermal effect. We note that black body radiation
present at $T_e\neq 0$ gives similar effect to the one of radiation generated by the axion. 
It will turns out that the axion effect is dominant over the thermal effect when $m_a>T$ ( not $T_e$ ) 
as long as real temperature $T$ is less than $100$mK,
when axion mass $\sim 10^{-5}$eV. When temperature is much larger than $100$mK, 
the energy distribution of electrons has no contribution of axion. So even for $m_a>T_e$, the sharp boundary is
located at $E_f+2T_e$ when real temperature is larger than $100$mK. 
But we must remember that 
it is unclear what is the value $T_e$ corresponding to real temperature $100$mK.  
The relation depends on each samples.

\vspace{0.1cm}
Now we explain how the width $\Delta B$ depends on axion mass and temperature.
We discuss it in the cases of $T_c>m_a$ and $T_c \le m_a$ separately. 
First, we start the case of $T_c>m_a$.

When the temperature $T_e$ is larger than the axion mass $m_a$, the energy distribution of electrons is almost identical to
the previous one with no axion effect. Thus, the width  $\Delta B$ is given in the way mentioned above.
Namely, the width $\Delta E_f$ leading to $\Delta B$ is given such that $\Delta E_f=2\delta_e(T_e)+4T_e$ for $T_e>m_a$.
The width decreases as $T_e$ decreases up to $T_c$ when $T_c>m_a$. At temperature $T_e=T_c$ ( $>m_a$ ),
it takes $2\delta+4T_c$. It further decreases as $T_e$ decreases up to $T_e=m_a$.
When $T_e \le m_a$, the width $\Delta E_f$ is given such that $\Delta E_f=2\delta+4m_a$ because
the energy distribution of electrons has sharp cutoff at $E_f+4m_a$. $\Delta E_f=2\delta+4m_a$ does not depend on $T_e$.
Thus,
the width saturates at $T_e=m_a$ when $T_c>m_a$. 
That is, $\Delta E_f$ is given in the following,

\begin{equation}
\Delta E_f=2\delta_e(T_e)+4T_e \quad \mbox{for} \quad T_e>m_a  \quad \mbox{and} \quad \Delta E_f=2\delta+4m_a 
\quad \mbox{for} \quad T_e \le m_a 
\end{equation}
in the Hall bar with $T_c>m_a$. It saturates at $T_e=m_a$ and takes the value $\Delta E_f=2\delta+4m_a $.
When we decrease the size of Hall bar ( increase $\delta$ ), the saturation temperature does not change, while the width $\Delta B$ 
corresponding to $\Delta E_f$ increases.

\begin{figure}[htp]
\centering
\includegraphics[width=0.8\hsize]{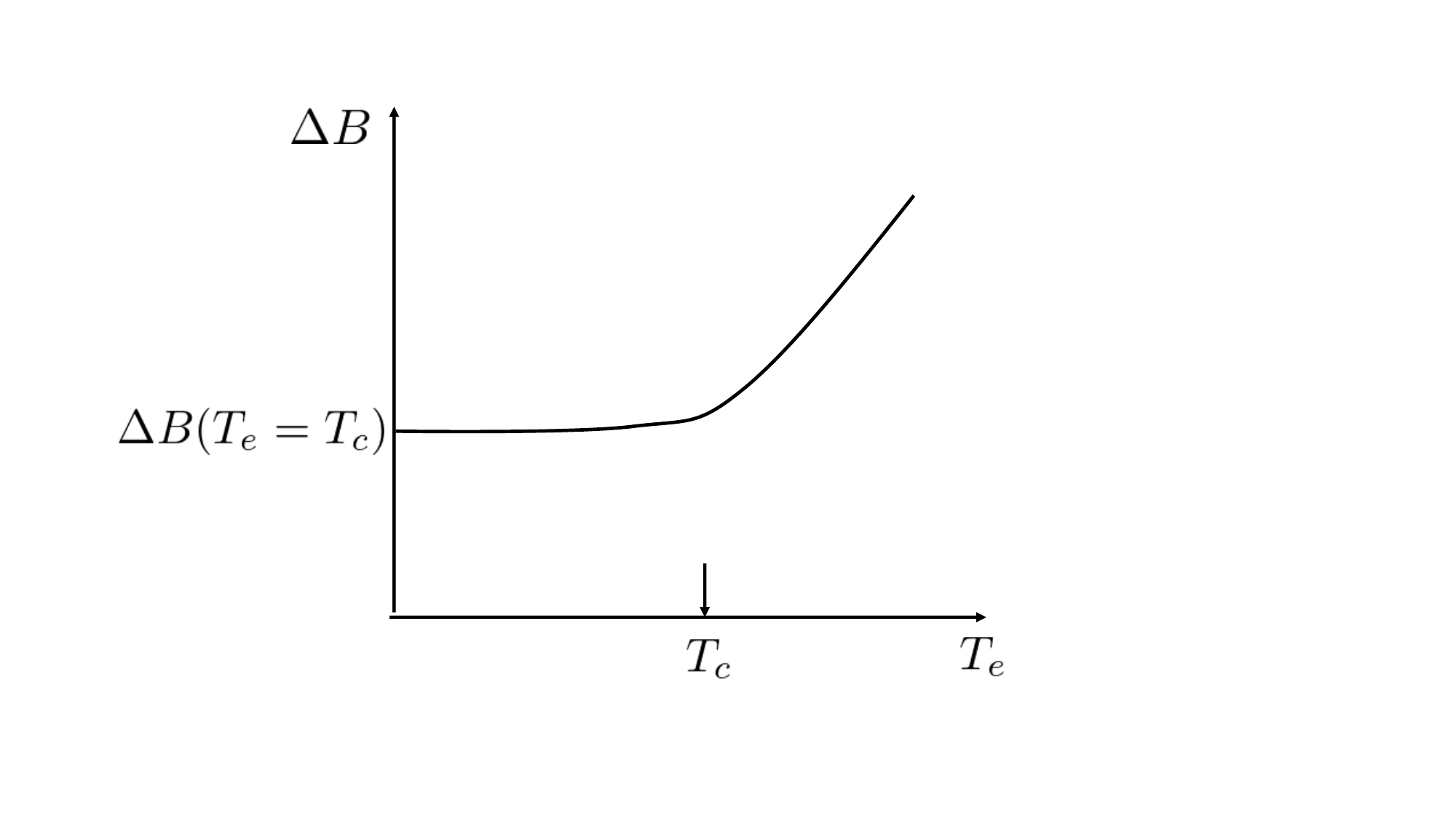}
\caption{Saturation of width $\Delta B$ in temperature when $T_c<m_a$}
\label{ee}
\end{figure}

\vspace{0.1cm}
Next, we discuss the case of $T_c \le m_a$. The width is given such that $\Delta E_f=2\delta_e(T_e)+4T_e$ for $T_e>m_a$, while
$\Delta E_f=2\delta_e(T_e)+4m_a$ for $T_c\le T_e\le m_a$. Obviously, it decreases with $T_e$, but saturates at $T_e=T_c$ 
because $\delta_e(T_c)=\delta$.
The width is given such that $\Delta E_f=2\delta+4m_a$ for $T_e \le T_c$.
That is, $\Delta E_f$ is given in the following,

\begin{equation}
\Delta E_f=2\delta_e(T_e)+4T_e \quad \mbox{for} \quad T_e>m_a  \quad \mbox{and} \quad \Delta E_f=2\delta_e(T_e)+4m_a 
\quad \mbox{for} \quad T_e\le m_a, 
\end{equation}
in the Hall bar with $T_c<m_a$. It saturates at $T_e=T_c<m_a$ and takes the value $ \Delta E_f=2\delta+4m_a $.
See Fig.\ref{ee}. The saturation temperature $T_e=T_c$ depends on the size of Hall bar.

\vspace{0.3cm}

To be summarized, the saturation temperature $T_s$ and the width of $\Delta E_f$ at $T_e\le T_s$ are 
given in the following. 
When $T_c>m_a$

\begin{equation}
\label{9}
T_s=m_a, \quad  \Delta E_f=2\delta+4m_a ,
\end{equation}

while, when $m_a>T_c$

\begin{equation}
\label{10}
T_s=T_c, \quad  \Delta E_f=2\delta+4m_a
\end{equation}

We should notice misunderstanding that
from the above formula $T_s=m_a$, the observation of the saturation temperature give real axion mass.
As we have stated, the temperature $T_e$ in the formula is not real temperature, although it is not far from real temperature $T$.
The formula $T_a=m_a$ in eq(\ref{9}) implies that the saturation temperature does not depend on the size of Hall bar.
On the other hand, the formula $T_s=T_c$ implies that the saturation temperature depends on the size of Hall bar.
In order to find the saturation temperature $T_s=m_a$, we must check the independence of $T_s$ on the size of Hall bar.  

Although the above result shows that the saturation temperature $T_s$ is equal to or less than the axion mass $m_a$ in any cases,
it is possible to have real saturation temperature large such as $1$K$\sim 10^{-4}$eV, which is larger than axion mass $\sim 10^{-5}$eV expected 
later in the present paper.

\vspace{0.1cm}

In the previous paper\cite{wanli},
it has been shown that as the size decreases, both of 
the saturation temperature $T_s=T_c$ and the width $\Delta B$ 
corresponding to $\Delta E_f$ increases. Indeed, it has been observed that $T_s \propto L_h^{-1}$.  
It indicates that $T_c \propto  L_h^{-1}$. It has been understood that the saturation is caused by finite size effect.
According to our analysis, such a feature of the saturation suggests
that the sample used in the paper has the feature $T_c<m_a$.
( The real saturation temperatures in the paper are in the range $300\rm mK \sim 10mK$. )

On the other hand, it has been shown in the reference\cite{sat5} 
that the saturation temperatures $T_s$ in samples with sizes $50\mu$m$\times200\mu$m, $200\mu$m$\times800\mu$m
and $800\mu$m$\times3200\mu$m
are almost identical ( real temperature $T\sim 30$mK ).  Obviously, finite size effect does not appear. 
Thus, it has been stated in the reference\cite{sat5} that the feature 
is caused by intrinsic decoherence. On the other hand, according to our analysis,
these samples may
have the feature of $T_c>m_a$ and the saturation temperature is given by the axion mass, $T_s=m_a$. 
Therefore, the saturation, in other words, the intrinsic decoherence observed in the paper is caused by the axion dark matter.

%

\vspace{0.1cm}

The paper\cite{sat5} suggests that the axion mass is given by the real saturation temperature $\simeq 30$mK.
Although the temperature is not identical to the axion mass $m_a$.
it suggests that it is near to the axion mass. 
Later we will find that the axion mass is strongly suggested to be equal to $(0.95\sim 0.99)\times 10^{-5}\mbox{eV}$
( $\sim 100$mK ) by experiments using microwaves imposed on Hall bar.

\section{external microwave and determination of axion mass}
\label{6}

We proceed to discuss the effect of external microwave imposed on Hall bar. We would like to know how the width $\Delta B$ ( or $\Delta E_f$ )
depends on the frequency $f$ of the microwave. Contrary to the observation of saturation temperature $T_s$,  
it will turn out that we can determine the axion mass by observing the saturation frequency $f_s$ of $\Delta B$.

\vspace{0.1cm}

The effect of the microwave is identical to the one of radiation generated by the axion. The difference is
that we can change frequency and power of the microwave. The power is in general much larger than the
power generated by the axion. The energy distribution of electrons is modified significantly, but
its power must be taken small enough not to increase the temperature of the Hall bar.
In actual experiments, the power has been taken sufficiently small for the temperature not to increase.  

We also simplify the effect of the radiations similarly to that of radiation generated by the axion.
That is, they modify energy distribution of electrons such that there is a sharp boundary in the distribution.
For instance it has sharp boundary at the energy $E_f+2\pi f$ when $2\pi f>T_e$. The states with energies larger than $E_f+2\pi f$ are empty. 
On the other hand, it has sharp boundary at the energy $E_f+2T_e$ when $2\pi f \le T_e$. 

\vspace{0.1cm}
We should make a comment that although the frequency $f$ used to define the sharp boundary is not real frequency,
the formula $f=m_a/2\pi$ obtained in subsequent discussions gives real axion mass $m_a$ by observing the real frequency $f$
in the experiment using external microwave. This is because both effects of radiations generated by axion and experimental apparatus
are identically simplified on energy distribution of electrons. Therefore, the formula $f=m_a/2\pi$ obtained later implies that 
the real frequency $f$ is equal to the real axion mass divided by $2\pi$.

\vspace{0.1cm}
We remind that the effect of external microwave on the width $\Delta B$ is very similar to the effect of temperature, as shown experimentally.
For example,  $\Delta B\propto T^{\kappa}$ for $T\to 0$ and $\Delta B\propto f^{\kappa}$ for $f \to 0$
with $\kappa= 0.4 \sim 0.7$ \cite{doo1}. The behavior is expected in scaling theory as a critical behavior. 
The external microwaves with frequency $f$ diminish phase coherent length just as thermal fluctuations in temperature $T$ diminish it
with correspondence $f \sim T$. 
Thus, we may take the effect of the external microwave on electrons in a similar way to the one of
temperature. 

The similarity can be understood by comparing the effect of the radiation with thermal effect.
Electrons absorb thermal energy $T$ and transit to states with higher energy, but they lose their energies emitting 
phonons. Similarly electrons absorb radiation energy $f$ and transit to states with higher energy, but they lose their energies emitting phonons. 
Only difference is that the thermal energy $E$ distributes around $T$, while radiation energy is given by a single frequency $f$. 
Thus, we expect that both effect of temperature and radiation is almost identical.

\vspace{0.1cm}
Here we should notice that external microwaves makes the coherent length of electrons decrease just as
temperature does. Thus, $\delta_e$ depends not only on temperature but also frequency of microwave such that
$\delta_e(T_e,f)$ increases as $T_e$ or $f$ increases with the condition $\delta_e(T_e,f=0)=\delta_e(T_e)$ and $\delta_e(T_e\le T_c)=\delta$.
It is naturally supposed that there is a critical frequency $f_c(T_e)$ such that $\delta_e(T_e,f)=\delta_e(T_e,f_c(T_e))$ for $f\le f_c(T_e)$. That is,
$\delta_e(T_e,f)$ does not decrease even more in $f$ for $f<f_c(T_e)$. It saturates at the critical frequency $f_c(T_e)$ when frequency $f$ decreases.
In other words, the energy region $E_{1+}+\delta_e(T_e,f)>E>E_{1+}-\delta_e(T_e,f)$ of the effectively extended states
does not change as long as $f<f_c(T_e)$.
The critical frequency $f_c(T_e)$ decreases with the decrease of the temperature $T_e$. Furthermore, for the temperature $T_e\le T_c$,  
$f_c(T_e)$ does not decrease less than $f_c(T_c)$. Namely, phase coherent length at temperature $T_e<T_c$ and 
frequency $f<f_c(T_e=T_c)$
is larger than the size of Hall bar. That is, $\delta_e(T_e,f_c(T_e))=\delta$ for $T_e\le T_c$.

Therefore, we speculate the dependence of $\delta_e(T_e,f)$ on temperature $T_e$ and frequency $f$ in the following,

\begin{eqnarray}
\delta_e(T_e,f=0)&=&\delta_e(T_e)=\delta_e(T_e,f)=\delta_e(T_e,f_c(T_e)) \quad \mbox{for $f \le f_c(T_e)$}, \quad  \delta_e(T_e,f_c(T_e))=\delta
\quad \mbox{for $T_e\le T_c$} \nonumber \\
f_c(T_e')&<&f_c(T_e) \quad \mbox{for $T_e' < T_e$}
\end{eqnarray}
Based on the speculation, we examine how the width $\Delta E_f$ leading to $\Delta B$ behaves 
depending on temperature $T_e$ and frequency $f$.

\vspace{0.1cm}
As discussed before,
we discuss two cases of Hall bar separately. One is Hall bar with $T_c>m_a$, that is, Hall bar with small size of length.
The other one is that with $T_c\le m_a$, that is, Hall bar with large size.

\vspace{0.1cm}
First we consider the case $T_c>m_a$. When high temperature $T_e>T_c>m_a$,
the energy distribution of electrons is modified by the microwave with frequency $f$ such that
its sharp boundary is located at $E_f+4\pi f$ for $2\pi f >T_e$ or $E_f+2T_e$ for $T_e \ge 2\pi f$.
Thus, the width $\Delta E_f$ is given in the following,

\begin{equation}
\Delta E_f=2\delta_e(T_e,f)+8\pi f \quad \mbox{for} \quad 2\pi f>T_e \quad \mbox{and}
\quad \Delta E_f=2\delta_e(T_e,f)+4T_e\quad \mbox{for} \quad 2\pi f\le T_e.
\end{equation}
 
We can see that the width $\Delta E_f$ decreases with the decrease of the frequency $f$, but it saturates at the frequency $f=T_e/2\pi$ when $2\pi f_c(T_e)>T_e$,
while it does at $f_s=f_c(T_e)$ when $2\pi f_c(T_e)\le T_e$. In both cases, $\Delta E_f=2\delta_e(T_e,f_c(T_e))+4T_e$.
Therefore, in the case $T_e>T_c>m_a$, both of saturation frequency $f=T_e/2\pi$ and $f_s=f_c(T_e)$ decrease with 
the decrease of the temperature $T_e$. 

\vspace{0.1cm}
On the other hand, in the case, low temperature $T_e<m_a$ ( $<T_c$ ), we have 

\begin{equation}
\label{t>m}
\Delta E_f=2\delta_e(T_e,f) +8\pi f \quad \mbox{for} \quad 2\pi f>m_a \quad \mbox{and} \quad
\Delta E_f=2\delta_e(T_e,f) +4m_a \quad \mbox{for} \quad 2\pi f \le m_a.
\end{equation}
 
When $f_c(T_c)>m_a/2\pi$, we have two cases,

\begin{equation}
\Delta E_f=2\delta_e(T_e,f) +8\pi f \quad \mbox{for} \quad f>f_c(T_c)\,\, (>m_a/2\pi ) \quad
\mbox{and} \quad \Delta E_f=2\delta +8\pi f \quad \mbox{for} \quad  f\le f_c(T_c).
\end{equation}
because $\delta_e(T_e,f)=\delta_e(T_e,f_c(T_c))=\delta $ for $f\le f_c(T_c)$.

The width $\Delta E_f=2\delta_e(T_e,f) +8\pi f$ saturates at $f_s=m_a/2\pi$ where
 $\Delta E_f=2\delta +4m_a$. While, when $f_c(T_c) \le m_a/2\pi$, according to the equation(\ref{t>m}),
the width $\Delta E_f$ saturates at
$f_s=f_c(T_c)\le m_a/2\pi$
where $\Delta E_f=2\delta +4m_a$. 
Therefore,
in the case $T_e<m_a$ ( $<T_c$ ), 
both saturation frequency $f_s=m_a/2\pi$ and $f_s=f_c(T_c)$ does not decrease with temperature $T_e$.

\vspace{0.1cm}
Furthermore, in the case, middle temperature  $m_a<T_e$ ( $<T_c$ ), we have 

\begin{equation}
\Delta E_f=2\delta_e(T_e,f) +8\pi f \quad \mbox{for} \quad 2\pi f>T_e \quad \mbox{and} \quad 
\Delta E_f=2\delta_e(T_e,f) +4T_e \quad \mbox{for} \quad 2\pi f \le T_e. 
\end{equation}

When  $f_c(T_c)>T_e/2\pi$, we have two cases, 

\begin{equation}
\Delta E_f=2\delta_e(T_e,f) +8\pi f \quad \mbox{for} \quad f>f_c(T_c)\,\,(>T_e/2\pi ) \quad
\mbox{and} \quad \Delta E_f=2\delta +8\pi f \quad \mbox{for} \quad f\le f_c(T_c) .
\end{equation}

The width $\Delta E_f=2\delta_e(T_e,f) +8\pi f$ saturates at $f_s=T_e/2\pi$
where $\Delta E_f=2\delta +4T_e$. While, when $f_c(T_c) \le T_e/2\pi$, 
the width $\Delta E_f$ saturates at $f_s=f_c(T_c)$
 where $\Delta E_f=2\delta +4T_e$. Therefore,
in the case $m_a<T_e$ ( $<T_c$ ), the saturation frequency $f_s=T_e/2\pi$ decrease, but $f_s=f_c(T_c)$ does not decrease with 
the decrease of the temperature.

\vspace{0.2cm}
Secondly, we consider the case $T_c<m_a $, i.e. Hall bar with large size.
Similarly to the above argument, we consider two cases separately, that is, the case of high temperature $T_e>m_a$ and 
the case of low temperature $T_e<m_a$.  Then, we find the following results.

\vspace{0.1cm}

For sufficiently high temperature $T_e>m_a>T_c$, we have

\begin{equation}
\Delta E_f=2\delta_e(T_e,f) +8\pi f \quad \mbox{for} \quad 2\pi f>T_e  \quad \mbox{or} \quad \Delta E_f=2\delta_e(T_e,f) +4T_e 
\quad \mbox{for} \quad 2\pi f \le T_e, 
\end{equation}
When $f_c(T_e)>T_e/2\pi$, it saturates at the frequency $f_s=T_e/2\pi$ where $\Delta E_f=2\delta_e(T_e,f_c(T_e)) +4T_e$.
While, when $f_c(T_e)\le T_e/2\pi$, it saturates at $f_s=f_c(T_e)$ where $\Delta E_f=2\delta_e(T_e,f_c(T_e)) +4T_e$.
Therefore, in the case $T_e>m_a>T_c$, both of saturation frequencies $f_s=T_e/2\pi$ and $f_s=f_c(T_e)$ decrease 
with the decrease of the temperature $T_e$.
Similarly, the width $\Delta E_f=2\delta_e(T_e,f_c(T_e)) +4T_e$ at saturation frequencies
decreases with the decrease of $T_e$.

The saturation frequency $f_s=f_c(T_e)$ is possibly less than $m_a/2\pi$, while $f_s=T_e/2\pi$ is larger than $m_a/2\pi$.
Indeed, the cases with the saturation frequency $f_s=f_c(T_e)<1$GHz have been observed in previous papers\cite{balaban, hohls},
in which temperature is much larger than $100$mK. It may correspond to the case of the saturation frequency $f_s=f_c(T_e)<m_a/2\pi $ or
$f_s=T_e/2$. Furthermore, the reference\cite{balaban} shows that $f_s\propto T_e$.

\vspace{0.1cm}

For low temperature $ T_e<m_a $, but $T_c<T_e$ ( that is, $T_c<T_e<m_a$ ), we have

\begin{equation}
\Delta E_f=2\delta_e(T_e,f) +8\pi f \quad \mbox{for} \quad 2\pi f>m_a  \quad \mbox{or} \quad \Delta E_f=2\delta_e(T_e,f) +4m_a 
\quad \mbox{for} \quad 2\pi f \le m_a 
\end{equation}
When $f_c(T_e)>m_a/2\pi$, it saturates $f_s=m_a/2\pi$ where the width
$\Delta E_f=2\delta_e(T_e,f_c(T_e)) +4m_a$ deceases with the decrease of $T_e$.
While, when $f_c(T_e)\le m_a/2\pi$, it saturates at $f_s=f_c(T_e)$ where the width $\Delta E_f=2\delta_e(T_e,f_c(T_e)) +4m_a$ decreases
as the temperature decreases up to $T_e=T_c$.

\vspace{0.1cm}

On the other hand, for sufficiently low temperature $T_e<T_c$ ( i.e. $T_e<T_c<m_a$ ), when $f_c(T_c)>m_a/2\pi$,
we find that the width $\Delta E_f$ saturates at $f_s=m_a/2\pi$ where $\Delta E_f=2\delta+4m_a$
because $\delta_e(T_e,m_a/2\pi)=\delta$, while
when $f_c(T_c) \le m_a/2\pi $,  it saturates at $f_s=f_c(T_c)$. In both cases, $\Delta E_f=2\delta+4m_a$ at the saturation.
Therefore, the width $\Delta B$ corresponding to $\Delta E_f=2\delta+4m_a$ as well as the saturation frequency $f_s=m_a/2\pi$ or $f_s=f_c(T_c)$
does not decrease with $T_e$.

\vspace{0.3cm}

To be summarized,

when $T_c>m_a$

\vspace{0.1cm}
\hspace{1cm} for $T_e>T_c>m_a$,
\begin{equation}
\label{17}
f_s=\frac{T_e}{2\pi} \quad \mbox{or} \quad f_s=f_c(T_e)\le \frac{T_e}{2\pi}, \quad \Delta E_f=2\delta_e(T_e,f_c(T_e))+4T_e
\end{equation}

\hspace{1cm} for $T_c>T_e>m_a$,
\begin{equation}
\label{18}
f_s=\frac{T_e}{2\pi} \quad \mbox{or} \quad f_s=f_c(T_c)\le \frac{T_e}{2\pi}, \quad \Delta E_f=2\delta+4T_e
\end{equation}

\hspace{1cm} for $T_c>m_a>T_e$,
\begin{equation}
\label{19}
f_s=\frac{m_a}{2\pi} \quad \mbox{or} \quad f_s=f_c(T_c)\le \frac{m_a}{2\pi}, \quad \Delta E_f=2\delta+4m_a
\end{equation}

when $m_a>T_c$,
\vspace{0.1cm}

\hspace{1cm} for $ T_e>m_a>T_c$, 
\begin{equation}
\label{20}
f_s=\frac{T_e}{2\pi} \quad \mbox{or} \quad f_s=f_c(T_e)\le\frac{T_e}{2\pi}, \quad \Delta E_f=2\delta_e(T_e,f_c(T_e))+4T_e
\end{equation}

\hspace{1cm} for $m_a>T_e>T_c$,
\begin{equation}
\label{21}
f_s=\frac{m_a}{2\pi} \quad \mbox{or} \quad f_s=f_c(T_e)\le \frac{m_a}{2\pi}, \quad \Delta E_f=2\delta_e(T_e,f_c(T_e))+4m_a
\end{equation}

\hspace{1cm} for $m_a>T_c>T_e$,
\begin{equation}
\label{22}
f_s=\frac{m_a}{2\pi} \quad \mbox{or} \quad f_s=f_c(T_c)\le \frac{m_a}{2\pi}, \quad \Delta E_f=2\delta+4m_a
\end{equation}

\begin{figure}[htp]
\centering
\includegraphics[width=0.8\hsize]{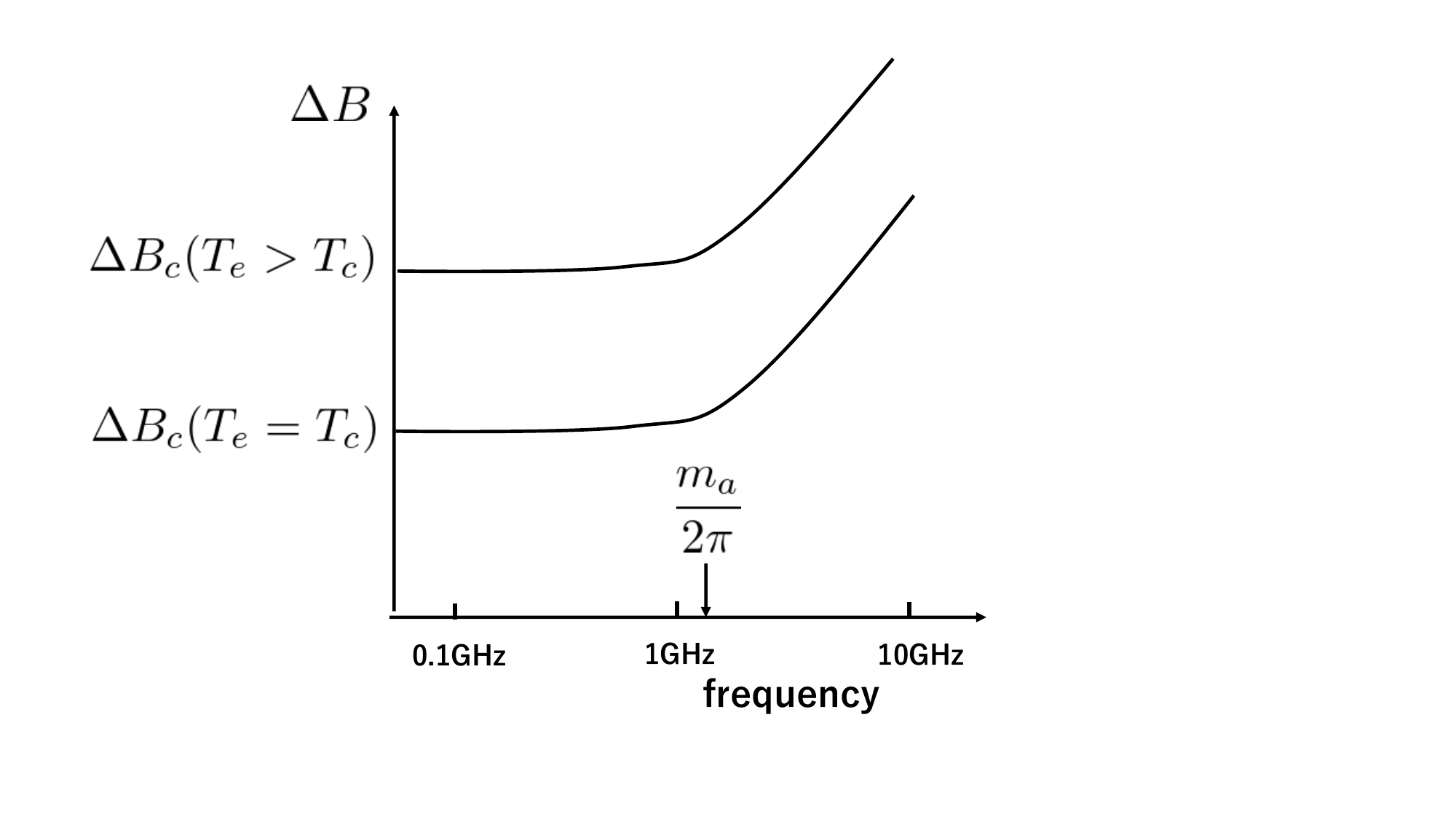}
\caption{Saturation of width $\Delta B$ at the frequency $f_s=m_a/2\pi$ for large Hall bar with $T_c<m_a$}
\label{ff}
\end{figure}

According to these results, we find how we determine the axion mass by imposing microwaves on Hall bar.
By searching the critical frequency $f_s=m_a/2\pi$, we can determine the axion mass. First,
we need low temperature such as $m_a>T_e$. Probably, it would be sufficient that the temperature is less than $50$mK. 
This can be expected from the reference\cite{sat5} we have mentioned in the section(\ref{5}).
The reference shows that the independence of the saturation temperature $T_s=m_a$ on the size of Hall bar is realized
at low temperature $30$mK.
Furthermore, there are two conditions which must be satisfied for critical frequency $f_s$ to be equal to $m_a/2\pi$.
The first condition is that the frequency $f_s$ does not decreases with the decrease of temperature.
The condition is satisfied by both $f_s=m_a/2\pi$ and $f_s=f_c(T_c)$.
Second one is that the frequency $f_s$ does not change with the change of the size of Hall bar. 
The condition is only satisfied by $f_s=m_a/2\pi$. In this way we can determine the axion mass
with experiment of quantum Hall effect using microwaves. In Fig.\ref{ff}, we schematically depict the curves 
of the width $\Delta B$ with temperatures $T_e=T_c$ and $T_e>T_c$ corresponding to the cases of equations (\ref{21}) and (\ref{22}).

\vspace{0.1cm}
There have been several experiments using microwaves. Among them, 
the reference\cite{balaban} shows that saturation frequency$f_s$ is proportional to temperature, i.e. $f_s\propto T_e$.
The frequency $f_s$ varies roughly from $2.3$GHz to $0.7$GHz corresponding to 
the temperatures $700$mK, $330$mK, and $150$mK. 
The temperatures may be larger than the axion mass $m_a$. Their widths $\Delta B$ also decrease with the decrease of the temperature.
The data correspond to either of the case in eq(\ref{17}), eq(\ref{18}) or eq(\ref{20}). 	
That is, $f_s=T_e/2\pi$ and $\Delta E_f=2\delta_e(T_e,f_c(T_e))+4T_e$ or $\Delta E_f=2\delta+4T_e$.

It appears in the reference\cite{engel} that saturation frequency $f_s\sim 2$GHz does not change for
temperatures $50$mK, $206$mK and $470$mK, although the reference involves ambiguities due to few data points.
Furthermore, it shows that $\Delta B$ decreases with the decrease of temperature. These data 
would correspond to the case in eq(\ref{21}). That is, $f_s=m_a/2\pi$ and $\Delta E_f=2\delta_e(T_e,f_c(T_e))+4m_a$ with $m_a>T_e>T_c$. 
The experiment suggests the axion mass $m_a\sim 10^{-5}$eV.

\vspace{0.1cm}
It is notable that 
an experiment\cite{saeed, doo,doo1} performed in temperature $35$mK has fine resolution of the frequency.
It has shown that the saturation arises at the frequency $f_s=2.3\mbox{GHz}\sim 2.4$GHz. 
Furthermore, almost identical frequency is observed in two samples with different sizes.
One has length $5.5$mm, while the other one has $20$mm both with the identical width $30\mu$m. 
Additionally, it has been stated in the reference\cite{saeed,doo,doo1} that
the temperature $35$mK is sufficiently low for temperature dependent effect to be negligible.
It might imply that small change of the temperature does not cause the change of their result. 
Therefore, the saturation frequency $f_s=2.3\mbox{GHz}\sim 2.4$GHz satisfies two conditions mentioned above.
The experiment corresponds to the case in eq(\ref{19}) or eq(\ref{22}). That is, $f_s=m_a/2\pi$ and $\Delta E_f=2\delta+4m_a$.
It also show the result consistent with $\Delta E_f=2\delta+4m_a$. The width $\Delta B$ corresponding to $\Delta E_f$
increases with the decrease of the Hall bar size.
The experiment strongly suggests that the mass of the dark matter axion is given by $m_a=(0.95\sim 0.99)\times 10^{-5}$eV.

\vspace{0.1cm}
It is interesting to see our previous prediction of axion mass ( $\sim 10^{-5}$eV ) derived with our
model\cite{axionstar,axionstar1} of fast radio burst. The fast radio burst is radio burst observed 
with various frequency range $0.2\mbox{GHz}\sim 8$GHz
and its energy is extremely large of the order of $10^{43}$GeV. 
The duration of the fast radio burst is short such as $\sim$millisecond or less. The fast radio burst arrives from
far Universe.  
Our model is that the burst is 
generated from the collision of axion star with neutron star or magnetized accretion disk of black hole.
The axion star is a composite of gravitationally bound axions. The radiations are generated by the axion star 
under strong magnetic field $\sim 10^{12}$G of neutron star or magnetized accretion disk. 
Because ionized gases emitting the radiations are high temperature, the line spectrum is affected by Doppler broadening so that 
they have wide range of frequencies. 
They also
receive gravitational red shift or red shift due to expanding Universe. Furthermore, they receive 
red or blue shift due to the rapid rotation of the accretion disk around black hole. These effects
make the line spectrum $m_a/2\pi$ generate the wide spectrum $0.2\mbox{GHz}\sim 8$GHz.
We have predicted by analyzing spectra of fast radio burst in our model that the axion mass is roughly equal to $10^{-5}$eV, which is 
coincident with the above result.

\section{axion domination over thermal noise}
\label{7}

We have assumed in the previous discussion that the energy distribution of electrons with energies larger than 
temperature $T$ is determined by axion effect when $m_a>T$. In this section we would like to 
confirm it by estimating energy power in Hall bar generated by the axion dark matter.  
We compute its signal noise ratio. The noise arises owing to black body radiation.
It turns out that the axion effect dominates over thermal noise when $m_a>T$ and the temperature
is less than $100$mK.

\vspace{0.1cm}
The axion dark matter generates radiations under strong magnetic field $B$.
Two dimensional electrons in Hall bar absorb the radiations and transit from localized states to effectively extended states.
The electrons in the effectively extended states carry electric currents so that Hall conductivity increases when such transitions arise.   
We discuss a transition from an electron in a state with energy $E_{\alpha}$ less than $E_f$ to the state with energy $E_{\alpha}+m_a$ ( $>E_{1+}-\delta$ ).
The wave function of a state with energy $E_{\alpha}$ in a Landau level is given by,

\begin{equation}
\Phi_{\alpha}=\exp(-iE_{\alpha}t)\int dk D_nf_{\alpha}(k)H_n\Big(\frac{x-l_B^2k}{l_B}\Big)\exp\Big(-\frac{(x-l_B^2k)^2}{2l_B^2}\Big)\exp(-iky),
\end{equation}
with $D_n=(2^{n+1}n!\pi^{3/2}l_B)^{-1/2}$, where  $H_n(x)$ denotes Hermite polynomials. 
The momentum $k$ characterizes degenerate state in a Landau level without disorder potential $V$.
The function $f_{\alpha}(k)$ is taken such that the wave function $\Phi_{\alpha}$ is an eigenstate of electron's Hamiltonian with potential $V$.
Normalization is such that that $1=\int dxdy \overline{\Phi_{\alpha}(x,y)}\Phi_{\alpha}(x,y)=\int dk \overline {f_{\alpha}(k)}f_{\alpha}(k)$.
The state is characterized by index $\alpha$. Here we assume no mixing between different Landau levels or different spin states.
The disorder potential $V$ is supposed to be much less than the energy difference between different Landau levels or different spin states.

Electron in the state $\alpha$ transits to a state $\beta$ with energy $E_{\alpha}+m_a$ by absorbing radiation. 
%
The transition amplitude is proportional to 

\begin{equation}
<\beta |H_a|\alpha>=\int dxdy \overline{\Phi}_{\beta}(x,y)\frac{-ie\vec{A}_a\cdot\vec{P}}{m^{\ast}}\Phi_{\alpha}(x,y)=
i(E_{\beta}-E_{\alpha})e\vec{A}_a\cdot <\beta|\vec{x}|\alpha>\equiv i(E_{\beta}-E_{\alpha})e\vec{A}_a\cdot \vec{L}_{\alpha\beta}
\end{equation}
where  $\vec{L}_{\alpha\beta}$( $\equiv <\beta|\vec{x}|\alpha>$ ) 
denotes a length scale of overlapping region of two states $\Phi_{\alpha}$ and $\Phi_{\beta}$;
$\vec{L}_{\alpha\beta}\equiv \int dxdy \overline{\Phi}_{\beta}(x,y)\vec{x}\Phi_{\alpha}(x,y)$.
$\vec{A}_a=\vec{\epsilon}\, g_{a\gamma\gamma}a_0B/m_a$ with polarization vector $\vec{\epsilon}$ pointing 
in two dimensional plane of Hall bar with $|\vec{\epsilon}|=1$.

%
%

The number of electrons $N$ making transitions per unit time by absorbing the radiation $\vec{A}_a$
from states with energies $E_{\alpha}$ lower than 
$E_f$ ( $<E_{1+}-\delta$ ) to states with energies $E_{\beta}$ larger than $E_{1+}-\delta$ is given by 

\begin{equation}
\dot{N}=2\pi S^2\int dE_{\alpha}\rho({E_{\alpha}+m_a})\rho(E_{\alpha})m_a^2\Big(e\vec{A}_a\cdot \vec{L}_{\alpha\beta}\Big)^2
\end{equation}
with surface area $S$ of two dimensional electrons. $\rho(E)$ denotes density of state. We explicitly use the formula,

\begin{equation}
\rho(E)=\rho_0\sqrt{1-\Big(\frac{E-E_{1+}}{\Delta E}\Big)^2} \quad \mbox{with} \quad |E-E_{1+}|\le \Delta E 
\quad \mbox{otherwise} \quad \rho(E)=0
\end{equation} 
with $\rho_0=(eB/2\pi)\times 2/(\pi \Delta E)$,
where $\int dE \rho(E)=eB/2\pi$ represents the number density of electrons in a Landau level;
$\int dE \rho(E)\simeq 2.4\times 10^{11}\rm cm^{-2}$$(B/10T)$. The width $\Delta E$ \cite{width} is of the order of 
$\sim 10^{-4}$eV. Tentatively, we take $\delta=\Delta E/10\sim 10^{-5}$eV.
Here we consider the energy region of effectively extended states such that 
$E_{1+}+\delta>E>E_{1+}-\delta$, assuming zero temperature or much less than $T_c$; $\delta_e(T_e)=\delta$ for $T_e<T_c$.


\vspace{0.1cm}

We consider the transition of electrons $\alpha$ with the energy $E_{\alpha}$ ( $<E_f-\delta$ ) to the states $\beta$ 
with the energy $E_{\beta}$ in the range $E_{1+}-\delta<E_{\beta}<E_{1+}+\delta$. 
When $E_f<E_{1+}-\delta-m_a$, such a transition does not arise. So, Hall conductivity does not increase. But when Fermi energy $E_f$ increases and
take a value as $E_f=E_{1+}-\delta-m_a$, such a transition begins to arise. The Hall conductivity also begins to increase.
Further increase of Fermi energy makes the Hall conductivity larger. Eventually, the conductivity  
arrives at the next plateau when Fermi energy reaches $E_f=E_{1+}+\delta$.

\vspace{0.1cm}

For simplicity, we assume in the estimation of $\dot{N}$ that the length $L_{\alpha\beta}$ is independent of the states $\alpha$ and $\beta$;
$L_{\alpha\beta}=A\l_B$. 
Then, the energy power $P_a=m_a\dot{N}$ of the axion absorbed in Hall bar is given by

\begin{eqnarray}
\label{N}
P_a&=&\int dE_{\alpha}2\pi S^2\rho({E_{\alpha}+m_a})\rho(E_{\alpha})m_a^3
\Big(e\vec{A}_a\cdot \vec{L}_{\alpha\beta}\Big)^2
\simeq 2\pi S^2m_a^3\Big(\frac{eB}{2\pi}\frac{2}{\pi \Delta E}\Big)^2e^2A_0^2A^2l_B^2 \times \Delta, \nonumber \\
&\sim& 1.9\times 10^{-19}\mbox{W}\Big(\frac{A}{10^4}\Big)^2 \Big(\frac{S}{10^{-3}\mbox{cm$^2$}}\Big)^2
\Big(\frac{0.5\times10^{-4}\mbox{eV}}{\Delta E}\Big)^2
\Big(\frac{\rho_d}{0.3\,\mbox{GeVcm$^{-3}$}}\Big)\Big(\frac{B}{10\rm T}\Big)^3\Big(\frac{m_a}{10^{-5}\mbox{eV}}\Big)^2\Big(\frac{g_{\gamma}}{1.0}\Big)^2
\end{eqnarray}
with $\rho(E_{\alpha})=\rho_0\sqrt{1-((E_{\alpha}-E_{1+})/\Delta E)^2}\simeq \rho_0$ because of our choices $\delta/\Delta E\sim 1/10$
and $\Delta E=0.5\times 10^{-4}$eV.
We have put  $(e\vec{A}_a\cdot \vec{L}_{\alpha\beta})^2\equiv e^2A_0^2A^2l_B^2$ with $A_0=g_{a\gamma\gamma}a_0B/m_a$.
Thus, the integral is trivial $\int dE_{\alpha}\equiv \Delta$, in which
the integration $\int dE_{\alpha}$ is taken over the range 
$E_f \ge E_{\alpha}\ge E_{1+}-\delta-m_a $. $\Delta=E_f-(E_{1+}-\delta-m_a)$
for $E_{1+}+\delta -m_a \ge E_f$, otherwise $\Delta=2\delta$.
We have taken that Fermi energy $E_f=E_{1+}-\delta$  in the above estimation and so $\Delta=m_a =2\delta$ with 
our choice $\delta=\Delta E/10$. 
We have tentatively assumed that the surface area of two dimensional electrons $S=10^{-3}\rm cm^2$ 
and the length scale, $Al_B\sim 10^4l_B$ of the overlapping 
between localized states $\alpha$ and effective extended states. We remember $g_{\gamma}=0.37$ for DFSZ axion model
and $g_{\gamma}=-0.96$ for KSVZ axion model.

\vspace{0.1cm}
We compare it with thermal noise by taking identical Fermi energy $E_f=E_{1+}-\delta$. 
The energies $\omega$ of the black body radiations are approximately restricted to be smaller than the temperature $T$; $\omega<T$. 
Localized electrons with energies $E_{\alpha}$ less than Fermi energy $E_f$ can be transited to effective extended states by absorbing the
radiations only when $E_{\alpha}+\omega>E_{1+}-\delta$. These electrons contribute Hall conductance.
Thus, when the temperature is sufficiently large such as $T>E_{1+}-\delta -E_f$, 
the radiations with the energies $\omega$ in the range, $T>\omega>E_{1+}-\delta -E_f$, are absorbed 
and the Hall conductance increases. The energy power $P_{th}$ of the thermal noise is given by $P_{th}=T(T-(E_{1+}-\delta -E_f))/2\pi$.
On the other hand, when the temperature is less than $E_{1+}-\delta -E_f$, the black body radiations do not contribute
the increase of the Hall conductance. ( Even if localized electrons absorb the radiations, they are only transmitted to localized states, not to effective 
extended states. )

Therefore, the increase of the Hall conductance is only caused by the axion dark matter when temperature $T$ and axion mass $m_a$ satisfy
the condition $m_a>E_{1+}-\delta -E_f>T$. 
The thermal noise does not contribute to the Hall conductance. This is the case in the previous section that the width $\Delta E_f$ is given by
$\Delta E_f=2\delta +4m_a$. 

When the temperature increases larger than $E_{1+}-\delta -E_f$, black body radiations also contribute to the increase of the Hall conductance.
Then, we need to find which contribution is dominant, axion effect or thermal one.

\vspace{0.1cm}
We compare the energy power $P_a$ of axion with thermal noise, $P_{th}=T(T-(E_{1+}-\delta-E_f))/2\pi$. 
In order to do so, we take Fermi energy such as $E_f =E_{1+}-\delta$ which is identical to the one used in the estimation of  $P_a$ in eq(\ref{N}).

SN ratio is given in the following,

\begin{eqnarray}
&&\frac{P_a\sqrt{\big(\frac{T-(E_{1+}-\delta-E_f)}{2\pi}\big) \times 1\mbox{s}}}{P_{th}} \nonumber \\
&=& \frac{P_a\sqrt{2\pi T \times 1\mbox{s}}}{T^2}
\simeq 	3.0\times \Big(\frac{A}{10^4}\Big)^2\Big(\frac{S}{10^{-3}\mbox{cm$^2$}}\Big)^2\Big(\frac{0.5\times10^{-4}\mbox{eV}}{\Delta E}\Big)^2\Big(\frac{\rho_d}{0.3\,\mbox{GeVcm$^{-3}$}}\Big)\nonumber \\ 
&\times& \Big(\frac{B}{10\rm T}\Big)^3
\Big(\frac{m_a}{10^{-5}\rm eV}\Big)^2 \Big(\frac{100\mbox{mK}}{T}\Big)^{3/2}\Big(\frac{g_{\gamma}}{1.0}\Big)^2
\end{eqnarray}
with $T=100$mK ( $\simeq 8.6\times 10^{-6}$eV ) and $m_a=10^{-5}$eV ( $\simeq 116$mK ),
where we have assumed that it takes one second for each measurement of Hall conductance for given magnetic field $B$. 
The result holds for $m_a \le 2\delta$, otherwise, it is proportional to $m_a\delta$, not $m_a^2$.

\vspace{0.1cm}

The result of the SN ratio means the following. We have supposed in the estimation that $m_a>T$ and 
Fermi energy $E_f =E_{1+}-\delta$. There are electrons occupying effective extended states with energies $E$ 
( $E_{1+}+\delta>E>E_f=E_{1+}-\delta$ ). They are transited from localized states by thermal effect or axion effect.
Among them, electrons with energies larger than $E_f+T$ arise only due to the axion effect,
while the other ones
with energies $E$ ( $E_f+T \ge E \ge E_f$ ) arise due to both effects. 
The point is that the thermal effect does not produce electrons
with energies larger than $E_f+T$. This is because the black body radiations with energies $\omega$ ($ >T$ )
have been supposed to be suppressed in the estimation. 
Actually, as the estimation shows, the axion effect is dominant over the thermal effect when  
$m_a>T$ at least when temperature is less than $100$mK. The result is coincident with our previous discussion.
However,
actual effects of black body radiations does not vanish even for $\omega>T$ and it vanishes
exponentially $P_{th}\sim T^2(\omega/T)\exp(-\omega/T)$. So, the thermal effect for $\omega > T$ is still present although 
it is small. 
Indeed, thermal effect is present even for $\omega >T$ such that
it decreases smoothly for $116\mbox{mK}>\omega >100$mK. Because the width $\delta \omega$ ( =$116\rm mK-100$mK ) is small,
the dominance of the axion effect still holds. Obviously,
the thermal effect with $\omega \gg 100$mK vanishes exponentially. 
Therefore, as we have supposed, the thermal effect is not dominant for the phenomena 
with energy scale larger than $T$ as long as $m_a>T$ and $T\le100$mK.
Namely, when $m_a>T$, electrons with
energies larger than $E_f+T$ are transited from localized states only by the axion effect.
The argument has been used extensively in the above sections.


\vspace{0.1cm}
We should make a comment that the above result heavily depends on the assumption of $A=10^4$ (
$\vec{L}_{\alpha\beta}\equiv <\beta|\vec{x}|\alpha>=Al_B$ with $l_B\simeq 8.2\times 10^{-7}\mbox{cm}\sqrt{(10\mbox{T}/B)}$ ).
That is, the overlapping region between localized state $\alpha$ with energy $E_{\alpha}$ 
and extended state $\beta $ with energy $E_{\alpha}+m_a$ is comparable to the size of Hall bar. 
For instance, when the ratio of side lengths of a rectangular Hall bar is such as 
$4:1$ and its surface area $S=10^{-3}\rm cm^2$, the largest side length is about $\simeq 6.3\times 10^{-2}$cm.
Thus, the length of the overlapping region $Al_B\simeq 8.2\times 10^{-3}$cm is about $1/10$ of the largest side length. It implies that
the extended state $\beta$ has length equal to the size of Hall bar, while the localized state with energy by $m_a\sim 10^{-5}$eV less than
the energy of the extended state, has length scale $\sim 1/10$ of the side length of the Hall bar.  
If the overlapping region is smaller than $1/10$ of the side length, we need lower temperature than $100$mK for the axion effect to dominate
the thermal effect. 

\vspace{0.1cm}
We may examine the validity of the assumption in the following.
For instance, supposing $\delta=0.5\times 10^{-5}$eV ( $=m_a/2$ ),  
the minimum energy of the localized state $\alpha$ able to transit to extended state $\beta$
by absorbing radiation with energy $m_a=10^{-5}$eV
is given by $E_{\alpha}=E_{1+}-\delta -m_a=E_{1+}-1.5\times 10^{-5}$eV. 
According to the scaling formula $\xi(E)\to |E-E_{1+}|^{-2.4}$ for $E\to E_{1+}$,
of coherent length $\xi(E)$
of the state with energy $E$, we find that the ratio of the scale of the localized state $\alpha$ with energy $E_{\alpha}=E_{1+}-1.5\times 10^{-5}$eV
to that of extended state with energy $E_{\beta}=E_{1+}-\delta=E_{1+}-0.5\times 10^{-5}$eV is about 
$\xi(E_{\alpha})/\xi(E_{\beta})=(1.5/0.5)^{-2.4}\simeq 0.07$. Because the extension of
the extended state $\beta$
is equal to or larger than the physical size of Hall bar, the size of the localized state $\alpha$ is about $1/10 $ of the size of Hall bar. 
Therefore, the choice of $A=10^4$ in the above estimation is reasonable.

\vspace{0.1cm}
Anyway, in order for the axion effect to dominate over thermal effect, we need low temperature. Namely,
in order to find the axion mass in a way mentioned above, it is favorable to have large Hall bar with surface area $S$ like $10^{-3}\rm cm^2$
and low temperature less than $50$mK. Under these consideration, we understand 
that the experiment\cite{saeed,doo,doo1} is appropriate for the
search of axion, which  
suggests the axion mass $m_a\simeq 0.9\times 10^{-5}$eV. The device in the experiment is cooled down $35$mK and
two dimensional electrons have much larger surface area
than $10^{-3}\rm cm^2$. Indeed, a coplanar waveguide with length $20$mm$=2$cm and width $30\mu$m$=3\times 10^{-3}$cm is used in the experiment. 
So, $S=6\times 10^{-3}\rm cm^2$.

\section{confirmation of presence of axion dark matter} 
\label{8}

When we decrease temperature, the saturation of the width $\Delta B$ arises at a critical temperature $T_s=m_a$ or $T_s<T_c$( $\le m_a$ ).
As we have discussed, such a saturation is caused by the axion. Without the axion, the $\Delta B$ behaves such that
$\Delta B=\mbox{const.}+c\, T_s$ as $T_s\to 0$ ( c denotes a constant. )

In order to examine whether the saturation is really caused by the axion, we diminish the effect of the radiations 
produced by the axion.
In our previous paper\cite{iwa}, we have proposed the way of the examination. 
We use two parallel conducting slabs such that they are put parallel to magnetic field.   
Such conducting slabs shield the radiations produced outside the plates.
Furthermore, the conducting plates themselves produce radiations by the axion. 
But the electric fields of the radiations\cite{iwazaki01} produced between the slabs
are perpendicular to two dimensional electrons. Such radiations are not absorbed in the Hall bar because excitation energies 
of two dimensional electrons in the direction perpendicular to the Hall bar have much larger than the axion energy.
Therefore, we can make to diminish the axion effect using the slabs, and
we will see that the width $\Delta B$ may not saturate at a critical temperature;
it behaves such as $\Delta B=\mbox{const.}+c\, T_s$ as $T_s\to 0$.

\begin{figure}[htp]
\centering
\includegraphics[width=0.6\hsize]{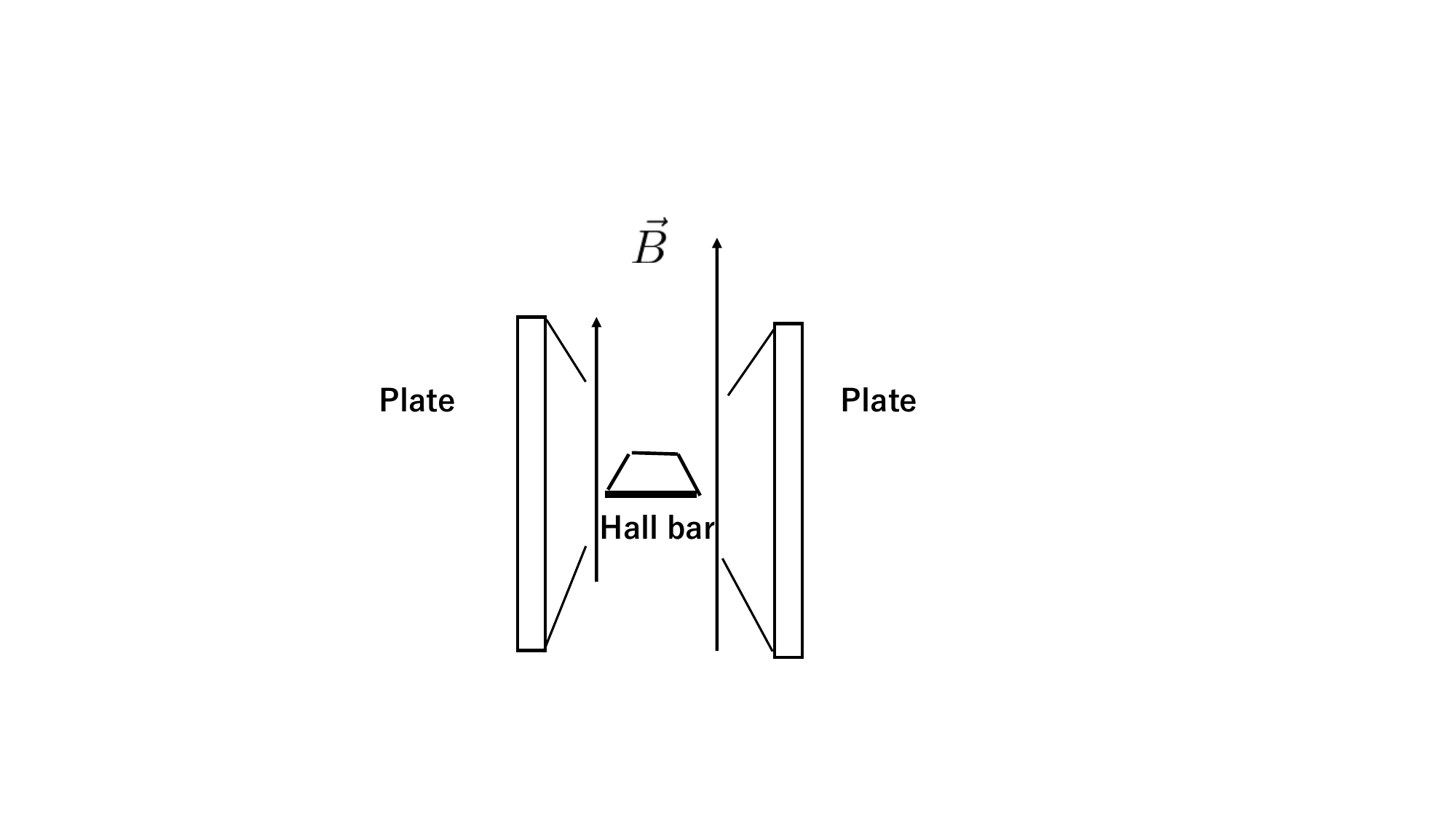}
\caption{Hall bar sandwiched by two conducting plates}
\label{d}
\end{figure}

\section{conclusion}
We have discussed the effect of the axion dark matter on integer quantum Hall effect. 
In particular, we have discussed the effect on plateau-plateau transition. The width $\Delta B$ between two plateaus decreases with the decrease of
temperature. But, it saturates at a critical temperature.
We have shown that the saturation takes place 
owing to the axion effect. Furthermore, similar saturation of the width arises in frequency 
when we impose microwaves.
By analyzing the axion effect on the saturation frequency $f_s$, we have presented two conditions which
$f_s$ must satisfy to give the formula $f_s=m_a/2\pi$.
The first condition is that $f_s$ does not change with the temperature. The second one is that
$f_s$ does not change with the size of Hall bar.
Such a saturation frequency should be studied in low temperature less than $50$mK with surface area of
Hall bar larger than $10^{-3}\rm cm^2$.
This is because the axion effect dominates over thermal noise in such low temperature and large
surface area.

\vspace{0.1cm}
It is remarkable to notice that such a saturation frequency has been obtained in 
previous experiment\cite{doo,doo1} with low temperature $\sim 35$mK and large sample such as $20$mm$\times 30\mu$m. 
The frequency has been obtained with high resolution of frequency.
Although it is not clear whether or not one of two conditions is satisfied,
it strongly suggests that the axion mass $m_a=(0.95\sim 0.99)\times 10^{-5}$eV.  
The condition not still confirmed is that $f_s$ does not change with temperature.
A similar frequency of saturation has been also 
observed in another experiment\cite{engel}, although the number of data sample is few and resolution of frequency is low.
 
Finally, we propose a way of the confirmation that the axion dark matter really causes the saturation.
Using two parallel conducting flat plates put parallel to magnetic field as shown in Fig.\ref{d}, 
we can diminish the axion effect on the two dimensional electrons. By diminishing the effect, we may observe
that the width $\Delta B$ does not saturate at a nonzero temperature.
It behaves such as $\Delta B=\mbox{const.}+c\, T_s$ as $T_s\to 0$.

\vspace{0.2cm}
The author
expresses thanks to A. Sawada for useful comments.
This work is supported in part by Grant-in-Aid for Scientific Research ( KAKENHI ), No.19K03832.



\end{document}